\documentclass[a4paper,12pt]{article}
\usepackage[latin1]{inputenc}
\usepackage[T1]{fontenc}
\usepackage{amsmath}
\usepackage{amsfonts}
\usepackage{amssymb}
\usepackage{color}
\usepackage{graphicx}
\usepackage{caption}
\usepackage{subcaption}
\usepackage{braket}
\usepackage{comment}

\newcommand{\be}{\begin{equation}}
\newcommand{\ee}{\end{equation}}
\newcommand{\ba}{\begin{eqnarray}}
\newcommand{\ea}{\end{eqnarray}}

\newcommand{\bea}{\begin{eqnarray}}
\newcommand{\eea}{\end{eqnarray}}

\begin{document}

\title{Multi-Skyrmions with orientational moduli}
\author{F. Canfora$^{1}$, G. Tallarita$^{2}$ \\
%EndAName
$^{1}$ {\normalsize \textit{Centro de Estudios Científicos (CECS), Casilla
1469, Valdivia, Chile.}}\\
$^{2}$ {\normalsize \textit{Departamento de Ciencias, Facultad de Artes
Liberales, Universidad Adolfo Ibáñez,} }\\
{\normalsize \textit{Santiago 7941169, Chile.}}\\
{\small canfora@cecs.cl, gianni.tallarita@uai.cl}}
\maketitle

\begin{abstract}
We analyze the mechanism of condensation of orientational moduli (as
introduced in \cite{shifman1}) on multi-Skyrmionic configurations of the
four-dimensional Skyrme model. The present analysis reveals interesting
novel features. First of all, the orientational moduli tend to decrease the
repulsive interactions between Skyrmions, the effect decreasing with the
increase of the Baryon number. Moreover, in the case of a single Skyrmion,
the appearance of moduli is energetically favorable if finite volume effects
are present. Otherwise, in the usual flat topologically trivial case, it is
not. In the low energy theory these solutions can be interpreted as
Skyrmions with additional isospin degrees of freedom.
\end{abstract}

\section{Introduction}

Skyrme's theory \cite{skyrme} is one of the most important models in
theoretical physics due to its wide range of applications in the low energy
limit of QCD (for a nice review, see \cite{manton}). Skyrme included his
famous term into the action of the four-dimensional non-linear sigma model
to allow the existence of static soliton solutions with finite energy,
called Skyrmions. A quite remarkable feature of this construction is that
the Skyrmion represents a Fermionic degree of freedom suitable to describe a
Baryon. This very deep intuition together with the close relation between
low-energy QCD and the Skyrme model itself were rigorously proved in the
classic papers \cite{48} \cite{49} \cite{50}. Moreover, from the
quantitative point of view, the theoretical and numerical computations based
on the Skyrme model are in good agreement with the experimental data in
nuclear physics \cite{51} \cite{52} \cite{53} \cite{54} \cite{55}.

Unfortunately, unlike what happens for instance in the case of monopoles and
instantons in Yang-Mills-Higgs theory (see \cite{manton}), it is extremely
difficult to construct exact non-trivial solutions of the Skyrme field
equations in which the non-linear effects of the Skyrme term are manifest.
The reason is that \textit{on flat topologically trivial space-times}, the
BPS bound on the energy in terms of the Baryon number (which was derived by
Skyrme himself), cannot be saturated for spherically symmetric Skyrmions.
Thus, until very recently, basically no exact non-trivial solution of the
four-dimensional Skyrme model was available. Even less exact results are
available in all the situations in which Skyrmions are confined to live
within a finite volume/ density (see, for instance, the analysis of \cite%
{klebanov}). In many of the applications of the Skyrme model in nuclear
physics and astrophysics one expects finite volume effects to be relevant,
but, of course, such effects usually make matters worse in terms of finding
exact solutions (as, for instance, finite volume effects break the
symmetries of the most common ansatz).

Using the formalism developed in \cite{41} \cite{46}, the first exact
solutions of the four-dimensional $SU(2)$ Skyrme model in which the
non-linear effects of the Skyrme term are manifest have been constructed in 
\cite{56}. Such solutions may possess non-trivial topological charges
different from the Baryon number. The conclusions of \cite{56} have been
strengthened in \cite{57}. Using these results, in \cite{58} the first exact
multi-Skyrmionic solutions of arbitrary Baryon number, living within a
finite volume and composed of interacting elementary $SU(2)$ Skyrmions, have
been constructed. The useful mathematical trick to construct these
configurations has been to place them within a suitable tube-shaped
finite-volume space-time (which, in the context of the Skyrme model, has
been also considered in \cite{58a}\ although with different motivations).
The main role of this topology is to maintain the finite-volume effects
without breaking the symmetries of the ansatz. The quite non-trivial
technical advantage of this choice to describe finite volume effects is that
it allows to compute explicitly the Skyrmion-Skyrmion interactions energy.
This computation shows clearly the repulsive character of Skyrmion-Skyrmion
interactions within this tube-shaped region. These results have been
generalized in \cite{58b}\ for the $SU(N)$ case. Moreover, these results
allowed the first analytic construction of gravitating Skyrmions in \cite%
{ACZ}.

This framework offers the intriguing possibility to analyze a very important
phenomenon which is typical of many topological defects: the condensation of
additional orientational moduli on their world-sheet (in the case of
Skyrmions, world-line). These orientational moduli are simply the Goldstone
bosons which arise by breaking a non-Abelian global symmetry on the core of
the original solitons. The corresponding solutions are called non-Abelian
solitons. The case of non-Abelian vortices has been of particular interest
as a possible description of the vortices responsible for the dual \cite{3}
confinement of quarks in QCD. Research in this direction, pioneered by \cite{Hanany:2003hp} \cite{Auzzi:2003fs} \cite{Shifman:2004dr} \cite{Hanany:2004ea}, led Shifman to
develop a particularly simple model (inspired by the famous Witten result on
superconducting cosmic strings \cite{wittenM}) in which orientational moduli
can condense on solitonic solutions \cite{shifman1}. The model was recently
used to construct the first case of non-Abelian vortices in holographic
models at strong coupling \cite{Tallarita:2015mca}. The analysis shows that
the main ingredients responsible for the presence of orientational moduli
are: a) a bulk theory with a global non-Abelian symmetry $G$ (which must be
unbroken initially) and which admits topological defects uncharged under
this symmetry (Skyrmions in the present case), b) the breaking of $G$ down
to a global subgroup $H$ on the given defect. The orientational moduli are
then the Goldstone bosons of the symmetry breaking pattern $G\rightarrow H$
and their low energy dynamics is described by the non-linear sigma model
with target space $G/H$.

In this paper we will adapt this model to produce Skyrmions with
orientational moduli. Given the previous results on multi-Skyrmion solutions
we are able to extend this study to the case of multi-solitonic
configurations. In particular this allows us to infer some characteristics
of the moduli interactions, of which little is known (see for example \cite%
{Kobayashi:2013axa}), and to study their finite volume physics. In fact,
most of the previously known applications of \cite{shifman1} dealt with
moduli localized on isolated topological soliton (for example see \cite{s1} 
\cite{p1} \cite{s2} \cite{p2}). This analysis reveals many interesting novel
features. First of all, the orientational moduli of \cite{shifman1}, when
finite volume effects are taken into account as in \cite{58}, tend to
decrease the repulsive interactions among Skyrmions which characterize the
multi-Skyrmionic configurations living in the tube-shaped regions mentioned
above. Moreover the appearance of moduli is energetically favorable if
finite volume effects are present. On the contrary, in the usual flat
topologically trivial case, we show that the non-Abelian Skyrmions are only
metastable solutions.\newline

The paper is organized as follows. In the second section, the general setup
in the usual flat metric with trivial topology is specified and the
numerical solutions are described. In the third section, the low energy
action for the orientational moduli is introduced. In the fourth section,
the setup to describe multi-Skyrmionic configurations at finite volume is
analyzed and the issue of moduli condensation is discussed. In the fifth
section some conclusions are drawn.

\section{Setup}

In order to consider orientational moduli on Skyrmion configurations, we
generalise the model introduced in \cite{shifman1}. We consider the action

\begin{equation}
S=\int d^{4}x\left( \mathcal{L}_{sk}+\mathcal{L}_{\chi }-\mathcal{L}%
_{int}\right)  \label{action}
\end{equation}%
where 
\begin{equation}
\mathcal{L}_{sk}=Tr\left( \frac{\kappa }{4}\left( U^{-1}\partial _{\mu
}U\right) ^{2}+\frac{\lambda }{4}\left[ U^{-1}\partial _{\mu
}U,U^{-1}\partial _{\nu }U\right] \left[ U^{-1}\partial ^{\mu
}U,U^{-1}\partial ^{\nu }U\right] \right) ,
\end{equation}%
\begin{equation}
\mathcal{L}_{chi}=\partial _{\mu }\chi ^{i}\partial ^{\mu }\chi ^{i},
\end{equation}%
\begin{equation}
\mathcal{L}_{int}=\gamma \left( Tr\left[ U+U^{-1}-2\right] +\Gamma \right)
\chi ^{i}\chi ^{i}-\beta (\chi ^{i}\chi ^{i})^{2}.
\end{equation}%
The Skyrme model possesses a conserved topological charge\footnote{%
Mathematically, this charge is the winding number associated to the third
homotopy class of maps from the three-dimensional sphere into $SU(2)$.}
which physically represents the Baryon number (see \cite{49} \cite{50}). Its
integral expression is%
\begin{equation}
W=-\frac{1}{24\pi ^{2}}\int_{\left\{ t=const\right\} }tr\left[ \left(
U^{-1}dU\right) ^{3}\right] \ ,  \label{windgen1}
\end{equation}%
where the integral is performed over $t=const$ hypersurfaces. Therefore,
when $W\neq 0$, the corresponding configuration cannot be deformed
continuously to the trivial vacua. \newline

In the action we recognise $\mathcal{L}_{sk}$ as the usual Skyrme Lagrangian
including the Skyrme term, $\mathcal{L}_{\chi}$ as the kinetic term of the
triplet $\chi^i$ charged under the global non-abelian group $O(3)$ and $%
\mathcal{L}_{int}$ as an interaction term. Just as per \cite{shifman1}, the
interaction term is designed to make the $\chi$ field condense on the core
of the Skyrmion. We will take the dimensionless parameter $0<\Gamma < 8$ and
all other parameters as positive. The mass dimensions of the fields and
parameters are $[U]=0$, $[\chi]=1$, $[\gamma]=[\kappa]=2$, $%
[\beta]=[\lambda]=0$. Our metric convention is $\eta_{\mu\nu}=(-,+,+,+)$ and
the convention on the group generators is $t_i = i\sigma_i$. We use a
spherical coordinate system and switch to the dimensionless units of $\rho = 
\sqrt{\kappa}r$ and $\tilde{\chi}= \chi/\sqrt{\kappa}$ (which we will adopt
from here throughout). \newline

We take the following ansatz for our fields 
\begin{eqnarray}
U &=&\cos (\alpha (\rho ))\mathbb{I}_{2}+\sin (\alpha (\rho ))n_{a}t_{a},
\label{spaccimma} \\
\chi ^{i} &=&\tilde{\chi}(\rho )(0,0,1),
\end{eqnarray}%
where $n_{a}=x_{a}/r$ and $\mathbb{I}_{2}$ is the $2$-dimensional identity
matrix. Then the energy functional reduces to

\begin{eqnarray}
\frac{E}{2\pi \sqrt{\kappa }} &=&\int_{0}^{\infty }d\rho \rho ^{2}\left( 
\frac{(\sin \alpha )^{2}}{\rho ^{2}}-\frac{2\lambda (\sin \alpha )^{2}\left(
-1+\cos (2\alpha )-4\rho ^{2}(\alpha ^{\prime 2})\right) }{\rho ^{4}}+\frac{1%
}{2}\alpha ^{\prime 2}\right.  \notag \\
&&\left. +\frac{\gamma }{\kappa }\left( -4+\Gamma +4\cos \alpha \right) 
\tilde{\chi}^{2}-\beta \tilde{\chi}^{4}+\tilde{\chi}^{\prime 2}\right) ,
\end{eqnarray}%
with $^{\prime }$ denoting differentiation with respect to $\rho $. It can
be easily checked that with the above ansatz for the Skyrmion and for the $%
\chi$ field the equations of motion coincide with the equations obtained as
stationary points of the above energy functional.

\subsection{Vacuum structure}

Assuming constant field profiles at infinity and minimizing the energy
functional at leading order in $(1/\rho)$ we obtain the following equations 
\begin{equation}
\tilde{\chi}_{vac}\sin\alpha_{vac}=0,
\end{equation}
\begin{equation}
\tilde{\chi}_{vac}\left(\tilde{\chi}_{vac}^2-\frac{\gamma}{2\beta\kappa}%
\left(-4+\Gamma+4\cos\alpha_{vac}\right)\right)=0.
\end{equation}

These set of equations have solutions of the form 
\begin{equation}
\tilde{\chi}_{vac}=0,\quad \alpha_{vac} = C,
\end{equation}
where $0 \leq C \leq 2\pi$ is an arbitrary angular constant, or 
\begin{equation}
\alpha_{vac} = n\pi, \quad \tilde{\chi}_{vac} = \pm \sqrt{\frac{\gamma}{%
2\beta\kappa}\left(-4+\Gamma+4(-1)^n\right)},
\end{equation}
where $n$ is an integer. Since we are working with the condition $0 < \Gamma
< 8$ (and all other parameters positive) then, for reality of $\tilde{\chi}%
_{vac}$ we must disregard $n$ odd and only consider even cases. For these
cases the corresponding energies of the vacuum solutions are 
\begin{equation}  \label{vac}
\frac{E_1}{2\pi\sqrt{\kappa}} =0 ,\quad \frac{E_2}{2\pi\sqrt{\kappa}} = 
\frac{\gamma^2\Gamma^2}{4|\beta|\kappa^2}.
\end{equation}
One can fix the constant $C=0$, meaning the true vacuum is at $U= \mathbb{I}%
_2$ and $\tilde{\chi}=0$. Note that the two branches coalesce at $\Gamma=0$.

\subsection{Energy minimization equations}

The equations which minimize the energy functional reduce to the coupled set
of ODE's

\begin{comment}
\begin{equation}
\alpha^{\prime \prime }+\frac{2\alpha^{\prime }}{r}-\frac{\sin(2\alpha)}{r^2}%
-\frac{8\lambda}{\kappa}\left[\frac{\sin(2\alpha)\sin(\alpha)^2}{r^4}-\frac{%
\sin(2\alpha)\alpha^{\prime 2}}{r^2}-\frac{2(\sin\alpha)^2\alpha^{\prime
\prime }}{r^2}\right]-\frac{4\gamma}{\kappa}\sin(\alpha)\chi^2=0,
\end{equation}
\begin{equation}
\chi^{\prime \prime }+\frac{2\chi^{\prime }}{r}-\gamma\left(-4+4\cos(%
\alpha)+\Gamma\right)\chi-2\beta\chi^3=0.
\end{equation}
\end{comment}

\begin{equation}
\alpha^{\prime \prime }+\frac{2\alpha^{\prime }}{\rho}-\frac{\sin(2\alpha)}{%
\rho^2}-8\lambda\left[\frac{\sin(2\alpha)\sin(\alpha)^2}{\rho^4}-\frac{%
\sin(2\alpha)\alpha^{\prime 2}}{\rho^2}-\frac{2(\sin\alpha)^2\alpha^{\prime
\prime }}{\rho^2}\right]-\frac{4\gamma}{\kappa}\sin(\alpha)\tilde{\chi}^2=0,
\end{equation}
\begin{equation}
\tilde{\chi}^{\prime \prime }+\frac{2\tilde{\chi}^{\prime }}{\rho}-\frac{%
\gamma}{\kappa}\left(-4+4\cos(\alpha)+\Gamma\right)\tilde{\chi}+2\beta\tilde{%
\chi}^3=0.
\end{equation}

The above equations must be solved numerically. In order to look for
Skyrmion solutions we solve these equations with the following boundary
conditions 
\begin{equation}
\alpha (0)=\pi ,\quad \tilde{\chi}^{\prime }(0)=0,
\end{equation}%
\begin{equation}
\alpha (\infty )=0,\quad \tilde{\chi}(\infty )=0.
\end{equation}

\begin{figure}[ptb]
\centering
\includegraphics[width=0.6\linewidth]{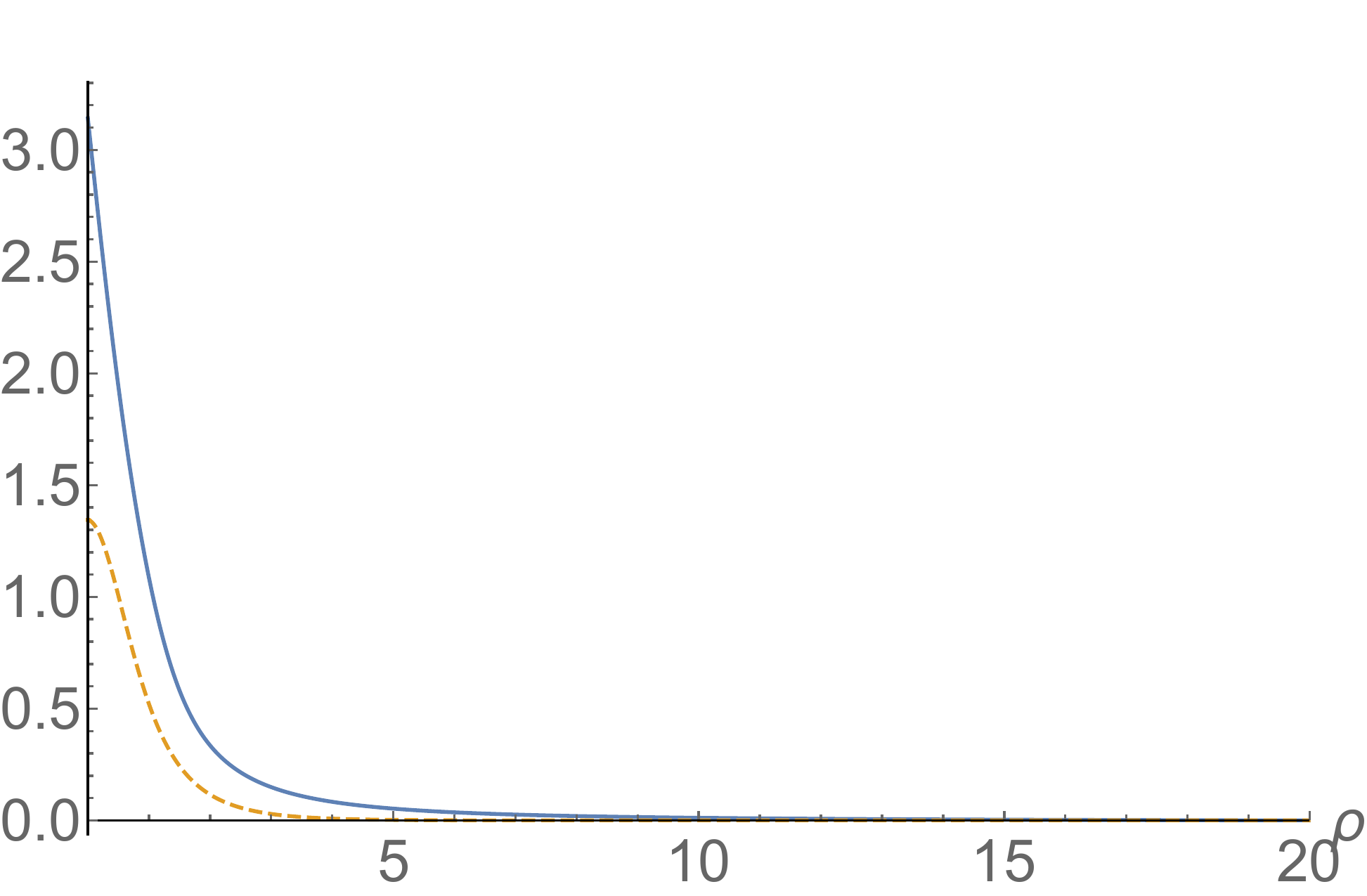}
\caption{Solution for $\protect\gamma/\protect\kappa= 1$, $\protect\beta%
=0.05 $, $\protect\lambda=1/8$ and $\Gamma=1$, solid line is $\protect\alpha$%
, dashed line is $\tilde{\protect\chi}$}
\label{fig4}
\end{figure}

Our numerical procedure is a second order finite difference procedure with
accuracy $\mathcal{O}(10^{-3})$. The solution is shown in figure \ref{fig4}.

\subsection{Mass}

In this section we will compare the mass of the solution with and without
moduli. This gives an indication as to which solution is energetically
preferred. The energy on our solution, with the parameters outlined in the
figure, evaluates to 
\begin{equation}
\frac{E}{2\pi\sqrt{\kappa}} = 5.966,
\end{equation}
whilst the energy of the $\tilde{\chi}=0$ solution is 
\begin{equation}
\frac{E}{2\pi\sqrt{\kappa}} = 5.804.
\end{equation}

Clearly the solution with moduli is higher in energy than the normal
Skyrmion. This result holds for the complete parameter range we have investigated for which our numerical procedure converged. In particular, the results for the masses diverge as $\protect\gamma/\protect\kappa$ and/or $\beta$ increases (for the parameters quoted in the solution, we found convergence up to $\beta = 0.09$ and $\gamma/\kappa = 2.625$). Both parameter changes increase the value of $\tilde{\chi}$ in the core.  Therefore, it must be an unstable, or at most metastable solution. This is in strong contrast to the case of vortices \cite{s1} or monopoles \cite{s2}, where the presence of additional moduli is energetically preferred.

In the next section we will argue that the solution is metastable.

\subsection{Stability}

Since the energy of the solution with additional moduli is higher than the
one without, we must check that the solution is not unstable. \newline

Perturbing our solutions with perturbations of the form 
\begin{equation}
\alpha =\alpha _{0}+\alpha _{p}e^{-i\omega t},\quad \chi^i=\left(\tilde{\chi}%
_{1p}e^{-i\omega t}, \tilde{\chi}_{2p}e^{-i\omega t},\tilde{\chi}_{0}+\tilde{%
\chi}_{3p}e^{-i\omega t}\right)
\end{equation}%
where $\alpha _{0}$ and $\tilde{\chi}_{0}$ are the background solutions
obtained above and $\tilde{\chi}_{ip}$ is a perturbation of the $\chi$ field
in the color direction $i$, we obtain the following set of coupled equations
for the perturbations 
\begin{equation}
\left( 
\begin{array}{cc}
\Delta _{1} & \Delta _{\chi } \\ 
\Delta _{\alpha } & \Delta _{2}%
\end{array}%
\right) \left( 
\begin{array}{c}
\alpha _{p} \\ 
\tilde{\chi}_{3p}%
\end{array}%
\right) =0,  \label{pertdef}
\end{equation}%
where, using the previously introduced dimensionless units and $\tilde{\omega%
}=\omega /\sqrt{\kappa }$, we have 
\begin{equation}
\Delta _{1}=V_{(1)}-\rho ^{2}\left( \left( \partial _{\rho }\Psi
_{(1)}\right) \partial _{\rho }+\Psi _{(1)}\partial _{\rho }^{2}\right) ,
\label{perdef1}
\end{equation}%
\begin{equation}
\Delta _{2}=V_{(2)}-\frac{1}{\rho }\left( \left( \partial _{\rho }\Psi
_{(2)}\right) \partial _{\rho }+\Psi _{(2)}\partial _{\rho }^{2}\right) \ ,
\label{pertdef2}
\end{equation}

and

\begin{eqnarray}
V_{(1)} &=&-\rho^{2}\left( \rho^{2}+8\lambda \right) \tilde{\omega}%
^2-8\lambda \cos \left( 4\alpha _{0}\right) +4\frac{\gamma }{\kappa}%
\rho^{4}\cos (\alpha _{0})\tilde{\chi} _{0}^{2}  \notag \\
&&+2\cos (2\alpha _{0})\left( 4\lambda +\rho^{2}(1 +4\lambda \tilde{\omega}
^{2})-8\rho^{2}\lambda (\alpha _{0}^{\prime 2}\right) -16\rho^{2}\lambda
\sin (2\alpha _{0})\alpha _{0}^{\prime \prime }\ ,
\end{eqnarray}%
\begin{equation}
V_{(2)}=\rho\left( \frac{\gamma}{\kappa} (-4+\Gamma )-\tilde{\omega} ^{2}+4%
\frac{\gamma}{\kappa}\cos (\alpha _{0})-6\beta \tilde{\chi _{0}}^{2}\right)\
,  \label{pertdef5}
\end{equation}
\begin{equation}
\Psi _{(1)}=(\rho^{2}+8\lambda (1-\cos (2\alpha _{0})))\ ,  \label{change1}
\end{equation}%
\begin{equation}
\Psi _{(2)}=\rho^{2}\ ,  \label{change2}
\end{equation}

\begin{equation}
\Delta _{\chi }=8\frac{\gamma }{\kappa}\rho^{4}\sin (\alpha _{0})\tilde{\chi}
_{0},  \label{pertdef3}
\end{equation}%
\begin{equation}
\Delta _{\alpha }=-4\frac{\gamma}{\kappa}\rho\sin (\alpha _{0})\tilde{\chi}
_{0}.  \label{pertdef4}
\end{equation}
\newline

The perturbations in the orthogonal colour directions decouple and obey the
following equation 
\begin{equation}
\tilde{\chi}_{ip}^{\prime \prime }+\frac{2}{\rho}\tilde{\chi}_{ip}^{\prime }+%
\tilde{\chi}_{ip}\left(\gamma(4-\Gamma-4\cos(\alpha_0))+\tilde{\omega}%
^2+2\beta\tilde{\chi}_0^2\right)=0,
\end{equation}
with $i=1,2$. We can put this equation in a convenient form by using the
transformation $\tilde{\chi}_{ip} = \hat{\chi}_{ip}/\rho$ to obtain the
Schr\"odinger equation 
\begin{equation}  \label{ortho}
\hat{\chi}^{\prime \prime }_{ip}+\hat{\chi}_{ip}\left(\gamma(4-\Gamma-4\cos(%
\alpha_0))+\tilde{\omega}^2+2\beta\tilde{\chi}_0^2\right)=0.
\end{equation}

In order to proceed with the stability analysis we make a change of
variables which simplifies the analysis of the coupled modes, we take%
\begin{eqnarray}
\alpha _{p} &=&Fu\ ,\ \ \tilde{\chi}_{p}=Gv\ ,  \label{change3} \\
F &=&\frac{1}{\sqrt{\Psi _{(1)}}}\ ,\ \ G=\frac{1}{\sqrt{\Psi _{(2)}}}\ ,
\label{change4}
\end{eqnarray}%
where $u$ and $v$ are the new unknowns in the linearized equations. In terms
of the new variables the linearized equations read:%
\begin{eqnarray}
\Delta _{\chi }Gv+V_{(1)}Fu-\rho ^{2}\left[ V_{(3)}u+\Psi _{(1)}F\partial
_{\rho }^{2}u\right] &=&0\ ,  \label{pertn1} \\
\Delta _{\alpha }Fu+V_{(2)}Gv-\frac{1}{\rho }\left[ V_{(4)}v+\Psi
_{(2)}G\partial _{\rho }^{2}v\right] &=&0\ ,  \label{pertn2}
\end{eqnarray}%
\begin{eqnarray}
V_{(3)} &=&\left[ \left( \partial _{\rho }\Psi _{(1)}\right) \left( \partial
_{\rho }F\right) +\Psi _{(1)}\partial _{\rho }^{2}F\right] \ ,
\label{change5} \\
V_{(4)} &=&\left[ \left( \partial _{\rho }\Psi _{(2)}\right) \left( \partial
_{\rho }G\right) +\Psi _{(2)}\partial _{\rho }^{2}G\right] \ .
\label{change6}
\end{eqnarray}%
Through this change of variables, the linearized system for the
perturbations equations (\ref{pertn1}) and (\ref{pertn2}) has been reduced
to a system of coupled Schr\"odinger equations. \newline

Our strategy to analyze the stability issue is the following: we find the
lowest energy normalizable solution of the coupled Schr\"odinger equations (%
\ref{change5})-(\ref{change6}). This solution will have normalizable
profiles for both $u$ and $v$ components and a minimum number of total
nodes. The solution is found by varying $\tilde{\omega}^{2}$. If $\tilde{%
\omega}$ is such that $\tilde{\omega}^{2}<0$ this indicates an instability
of the solution. We find numerically that $\tilde{\omega}^{2}\approx1.1>0$,
indicating that our solution is stable. For the perturbation modes in the
orthogonal colour directions an analogous analysis of equation (\ref{ortho})
yields $\tilde{\omega}^2 \approx 8>0$.

Clearly since the energy of our solution is higher than that with $\tilde{%
\chi}=0$ then our solution is actually metastable.

\section{Low energy theory of Orientational Moduli}

\label{sec3}

Standard quantization of the $SU(2)$ Skyrmion involves six moduli: three
translations of the Skyrmion centre and three space-flavour locked rotations
(see \cite{Shifman:2012zz} for a review). The quantum mechanical Hamiltonian
of these moduli is 
\begin{equation}
\mathcal{H}=M_{sk}+\frac{\vec{p}^{2}}{2M_{sk}}+\frac{1}{2I_{sk}}J(J+1),
\end{equation}%
where $J$ is a spin (or equivalently isospin) label. In the above $%
I_{sk}\approx 26\times\frac{51\pi \lambda ^{3/2}}{3\sqrt{\kappa }}$ \cite%
{Shifman:2012zz}. The presence of additional orientational moduli will give
an extra isospin quantum number. To see this we parametrise the moduli field 
$\tilde{\chi}$ as 
\begin{equation}
\chi^i=\tilde{\chi}(\rho )S^{i}(t),  \label{spaccimma2}
\end{equation}%
with $S^{i}$ a unit vector satisfying $S^{i}S^{i}=1$. Then inserting this
parametrisation into Eq.(\ref{action}) we obtain the low energy effective
action 
\begin{equation}
S_{le}=\frac{I_{1}}{2}\int dt\dot{S}^{i}\dot{S}^{i},\quad S^{i}S^{i}=1.
\end{equation}%
On symmetry grounds this is simply the $CP(1)$ non-linear sigma model which
results from the breaking pattern $O(3)\rightarrow U(1)$. The constant $%
I_{1} $, which plays the role of the moment of inertia, evaluates
numerically to 
\begin{equation}
\frac{I_{1}\sqrt{\kappa }}{4\pi }=\int_{0}^{\infty }\rho ^{2}\tilde{\chi}%
^{2}d\rho =0.372.
\end{equation}%
Quantization of this action follows the quantization of a rigid top (see 
\cite{s2}). It results in energy levels of the form 
\begin{equation}
E_{s}=\frac{1}{2I_{1}}s(s+1),
\end{equation}%
\ with degeneracy $2s+1$ and spherical harmonic eigenfunctions. \newline

As a curious possibility, in all the special cases in which%
\begin{equation}
\frac{I_{sk}}{I_{1}}=\frac{p}{q}\ ,\ \ p,q\in 
%TCIMACRO{\U{2115} }%
%BeginExpansion
\mathbb{N}
%EndExpansion
\ .  \label{accidegenr}
\end{equation}%
the eigenvalues of the Hamiltonian of the moduli read 
\begin{equation*}
\mathcal{H}=M_{sk}+\frac{\vec{p}^{2}}{2M_{sk}}+\frac{1}{2qI_{sk}}\left[
qJ(J+1)+ps(s+1)\right] \ .
\end{equation*}%
Therefore, in this case, the discrete part of the spectrum would be
determined by the integer number%
\begin{equation*}
N(J,s)=qJ(J+1)+ps(s+1)\ .
\end{equation*}%
It is easy to see that generically (once $p$ and $q$ are fixed) there is a
greater degeneracy of the energy states. Physically, an increase in
degeneracy of energy levels is associated with an enhancement of the
original symmetry. However, there is no obvious reason why eq. (\ref%
{accidegenr}) should be satisfied as this depends numerically on the
parameters of our system.

\section{Multi-Skyrmions in Tubular Topology}

One of the key properties of the hedgehog ansatz for $SU(2)$-valued scalar
fields is that it reduces a matrix system of coupled non-linear Partial
Differential Equations (PDEs) to a single scalar non-linear Ordinary
Differential Equation (ODE) for the soliton profile keeping alive, at the
same time, the corresponding topological features of the matrix-valued
field. This property is important in all contexts in which such ansatze are
used: from non-linear Sigma models to the Skyrme model (see \cite{manton}
for detailed reviews). Obviously, even when the scalar equation for the
soliton profile is not solvable analytically, the reduction of a matrix
system of coupled PDEs to a single ODE is a huge technical advantage both
theoretically and numerically. This property becomes even more important in
the cases in which multi-Skyrmionic configurations are considered. In \cite%
{58} and \cite{58b}, using the formalism developed in references \cite{41} 
\cite{46} \cite{56}, the matrix valued field equations of the
four-dimensional Skyrme sigma model have been reduced to a single scalar ODE
for the soliton profile in a sector of arbitrary topological charge (and in
such a way as to consider finite volume effects as well). These are the
basic technical results which we use here to construct non-Abelian
multi-Skyrmion configurations. \newline

Therefore, we change to a different topology with metric given by 
\begin{equation}
ds^{2}=-dt^{2}+dr^{2}+R^{2}(d\theta ^{2}+(\sin \theta )^{2}d\psi ^{2}).
\label{spaccimma3}
\end{equation}%

This is the metric for the Cartesian product space $\mathcal{R}^{(1,1)}\times S^2$. \newline

The usefulness of this geometry lies in the fact that it allows one to study
finite-volume effects (such effects usually make both numerical and
analytical studies very complicated: see \cite{klebanov}) keeping at the
same time the advantages coming from the symmetries of the hedgehog ansatz.
It describes three-dimensional cylinders whose sections are two-dimensional
spheres. Consequently, the parameter $R$ plays the role of the diameter of
the transverse sections of the tube. From the computational point of view,
the above choice of the metric is convenient since the radial variable $r$
which usually appears in front of the angles in the Minkowski metric is
replaced by a constant parameter (namely, $R$). This leads to considerable
simplifications, as the following analysis will prove. There is a price to
pay of course as the metric is curved. However, as the curvature of this
metric is proportional to $\frac{1}{R^{2}}$ one can easily consider a flat
limit by taking $%
R$ large. In the case of the Skyrme model, large means that $R$ should be
much larger than $1$ $fm$.

It is possible to be more precise about the effects of the local curvature
of the above metric. The Skyrme model is the leading approximation in the
large $\mathbf{N}$ limit of the low-energy action of QCD \cite{49}, however
subleading corrections in $1/\mathbf{N}$ do appear. Thus, from the practical
point of view, if one takes $R\sim 100\ fm$ the effects of the curvature are
already much smaller than other corrections to the Skyrme model arising from
QCD in the large $\mathbf{N}$\ expansion.

Therefore, the above metric can be considered as a \textquotedblleft
regulator\textquotedblright\. It is worth emphasizing that this
trick is also extremely useful when dealing with t'Hooft-Polyakov BPS
monopoles \cite{talla1}. We also make one other different consideration, in
this topology we work with values of $\beta <0$ for reasons which will soon
be explained. All other parameters are the same as before.

An important remark about the above geometry is in order. One may wonder
whether the usual identification of Skrmions with Fermions still holds on
the chosen curved geometry. In the reference \cite{curvedS1} the argument of 
\cite{curvedS2}\ has been generalized to curved orientable compact spaces.
As the spatial sections of the chosen geometry are orientable and compact,
we can claim that the third homotopy class is the Baryon charge in the
present case as well. This is a useful observation since, in the above
geometry, there are topological excitations in the Skyrme model with
topological charges different from the Baryon charge (see for instance \cite%
{56}). Such excitations with vanishing Baryon charge should be considered as
topological excitations of the Pionic sector. On the other hand, the
multi-Skyrmions which will be constructed here cannot decay into these
excitation of the Pionic sector as the Baryon charge is conserved.

In the new topology the energy (under the same ansatz) becomes

\begin{eqnarray}  \label{en2}
\frac{E}{2\pi\sqrt{\kappa}} = \int_0^{L} d\rho \left((\sin\alpha)^2-\frac{%
2\lambda(\sin\alpha)^2\left(-1+\cos(2\alpha)-4R^2\kappa(\alpha^{\prime
2})\right)}{\kappa R^2}+\frac{\kappa R^2}{2}(\alpha^{\prime 2 })\right. 
\notag \\
\left.+\gamma R^2\left(-4+\Gamma+4\cos\alpha\right)\tilde{\chi}^2+|\beta|
\kappa R^2\tilde{\chi}^4+\kappa R^2(\tilde{\chi}^{\prime 2})\right),
\end{eqnarray}

which is minimized by solving the equations

\begin{equation}  \label{eqR1}
\left( 1+\frac{16\lambda (\sin \alpha )^{2}}{\kappa R^{2}}\right) \alpha
^{\prime \prime }-\frac{\sin (2\alpha )}{\kappa R^{2}}\left( 1-8\lambda
\left( \alpha ^{\prime 2}-\frac{(\sin \alpha )^{2}}{\kappa R^{2}}\right)
\right) -\frac{4\gamma }{\kappa }(\sin \alpha )\tilde{\chi}^{2}=0,
\end{equation}%
\begin{equation}  \label{eqR2}
\frac{\gamma }{\kappa }\left( 4(-1+\cos \alpha )+\Gamma \right) \tilde{\chi}%
+2|\beta |\tilde{\chi}^{3}-\tilde{\chi}^{\prime \prime }=0.
\end{equation}%
Also in this case the full field equations of motion are equivalent to the
above coupled system of equations corresponding to the stationary condition
for the energy functional. \newline

Note that this finite volume topology has important consequences on the
energetic considerations. Firstly there is no origin in this geometry since $%
\rho=0$ is simply a convention of where to begin the length of the tube.
This implies that one can obtain energetically finite solutions without
vanishing derivatives there. Also, solutions have finite energy even though
they do not vanish at infinity, precisely because we consider a geometry
with finite length. In particular, since $\beta <0$ the lowest energy vacuum
solution of equation (\ref{en2}) has (see equation (\ref{vac}))

\begin{equation}
\alpha=(2n+1)\pi,\quad \tilde{\chi}_{vac} = \pm \sqrt{\frac{\gamma}{%
2|\beta|\kappa}\left(8-\Gamma\right)},
\end{equation}
with energy 
\begin{equation}
E_{vac} = -\frac{\gamma^2}{4|\beta|\kappa^2}(-8+\Gamma)^2.
\end{equation}

The important point is that solutions inside the finite geometry which do
not tend to the vacua at the extrema still have finite energy and,
consequently, still correspond to solutions with normalizable orientational
moduli, as we will shortly show. Therefore, in the remainder of this
section, when dealing with energetic considerations, we will not include
this constant shift in vacuum energy. \newline

For the Skyrme ansatz in eq. (\ref{spaccimma}) the Baryon number in eq. (\ref%
{windgen1}) reduces to%
\begin{equation*}
W=\frac{2}{\pi }\int_{\alpha (0)}^{\alpha (L)}\sin ^{2}\alpha d\alpha \ ,
\end{equation*}%
$L$ being the length of the tube. Thus, the Baryon number depends
exclusively on the boundary conditions for the Skyrmion profile $\alpha $.
In particular 
\begin{equation*}
\alpha (L)-\alpha (0)=n\pi \ \Rightarrow \ \ W=n\ .
\end{equation*}%
Actually, out of all the allowed integer values of $n$ (which denotes the number of Skyrmions), we can categorize the families of solutions between two possibilities (namely, $n$ even and $n$ odd)
which correspond to periodic and anti-periodic boundary conditions for the
Skyrme field $U$ in eq. (\ref{spaccimma}):%
\begin{eqnarray*}
n\ \ even\ &\Rightarrow &\ U(0)=U(L)\ , \\
n\ \ odd\ &\Rightarrow &\ U(0)=-U(L)\ .
\end{eqnarray*}

\subsection{Multi-Skyrmion solutions and energy considerations}

Here we present some solutions to the above equations. To find Skyrmion
solutions of Baryon number $n$ we use boundary conditions of the form (see
explanation above) 
\begin{equation}  \label{bc1}
\alpha(0)=0,\quad \tilde{\chi}^{\prime }(0)=0,
\end{equation}
\begin{equation}  \label{bc2}
\alpha(L)=n\pi,\quad \tilde{\chi}^{\prime }(L)=0,
\end{equation}
where $L$ is the dimensionless length of the tube we are placing the
Skyrmions in and $n$ the number of them. Note that we can choose Neumann
type boundary conditions for $\tilde{\chi}$ at both extremities of the
geometry since it is both bounded in length and has no origin. \newline

\begin{figure}[ptb]
\begin{subfigure}{.5\textwidth}
\centering
\includegraphics[width=0.8\linewidth]{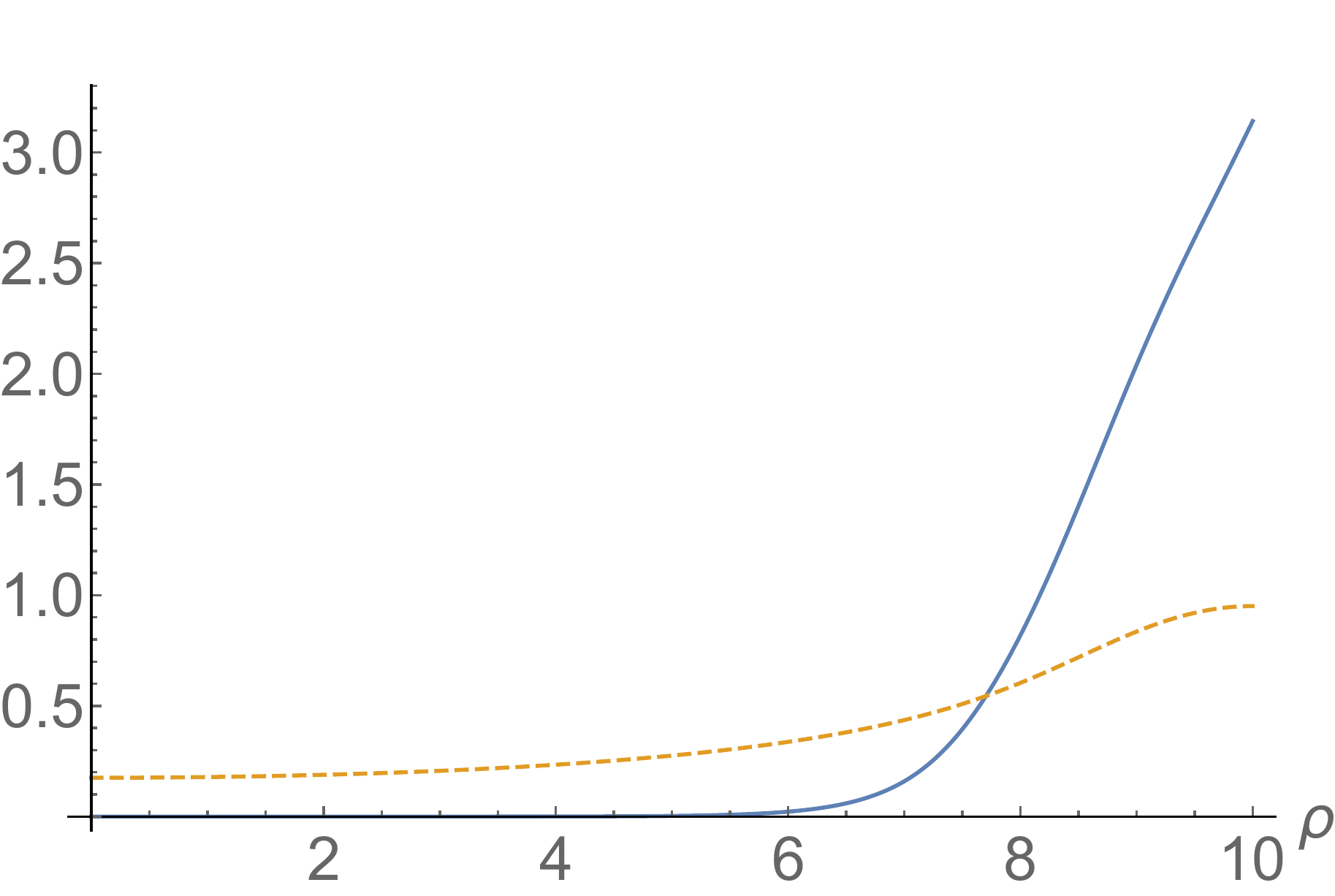}
\caption{}
\end{subfigure}
\begin{subfigure}{.5\textwidth}
\centering
\includegraphics[width=0.8\linewidth]{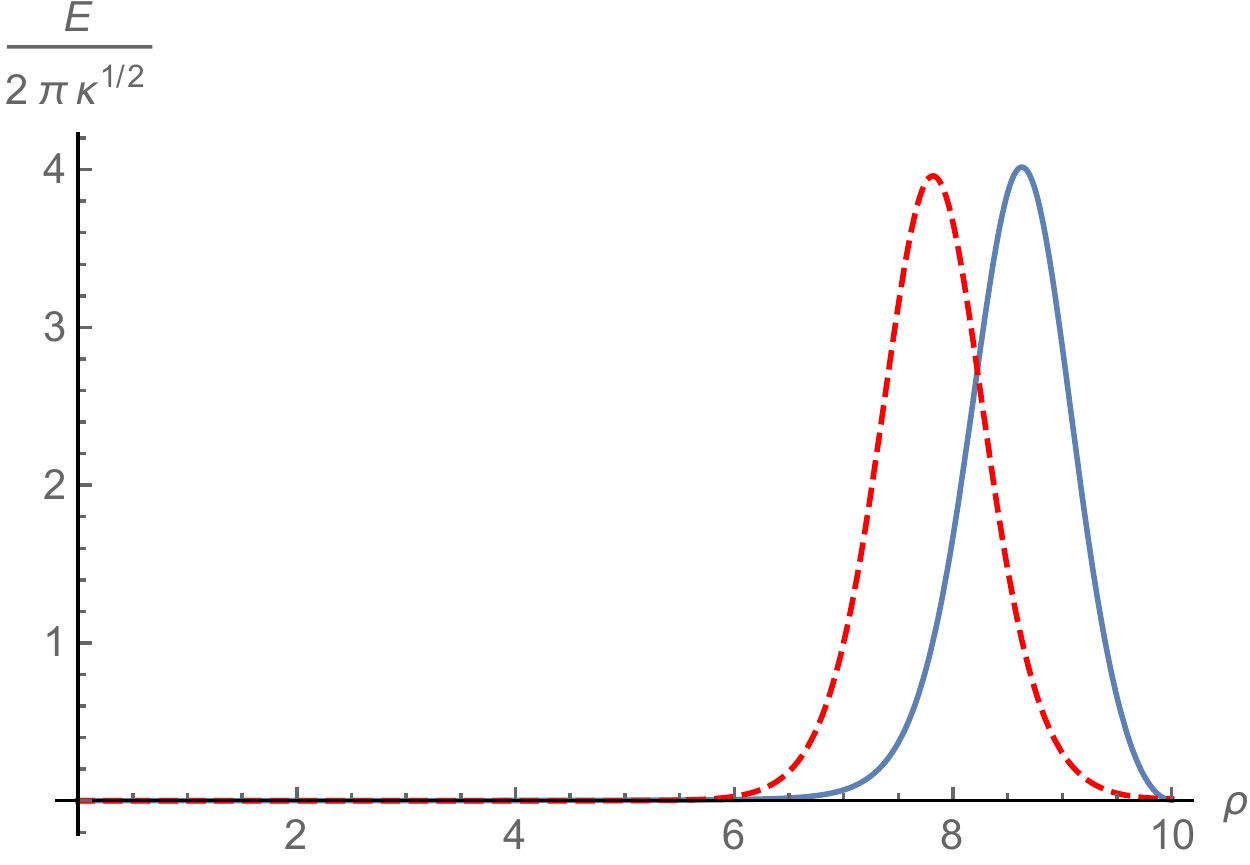}
\caption{}
\end{subfigure}
\caption{plot at $\protect\lambda = 1/8$, $|\protect\beta|=0.4$, $\Gamma=0$, 
$\protect\gamma/\protect\kappa =1/4$ and $\protect\sqrt{\protect\kappa}%
R=0.27 $. The solid line is $\protect\alpha$ and the dashed line is $\protect%
\chi$. In the energy plot we include in dashed red the plot for the energy
of the Skyrmion solution without any additional moduli. }
\label{fig12}
\end{figure}

\begin{figure}[ptb]
\begin{subfigure}{.5\textwidth}
\centering
\includegraphics[width=0.8\linewidth]{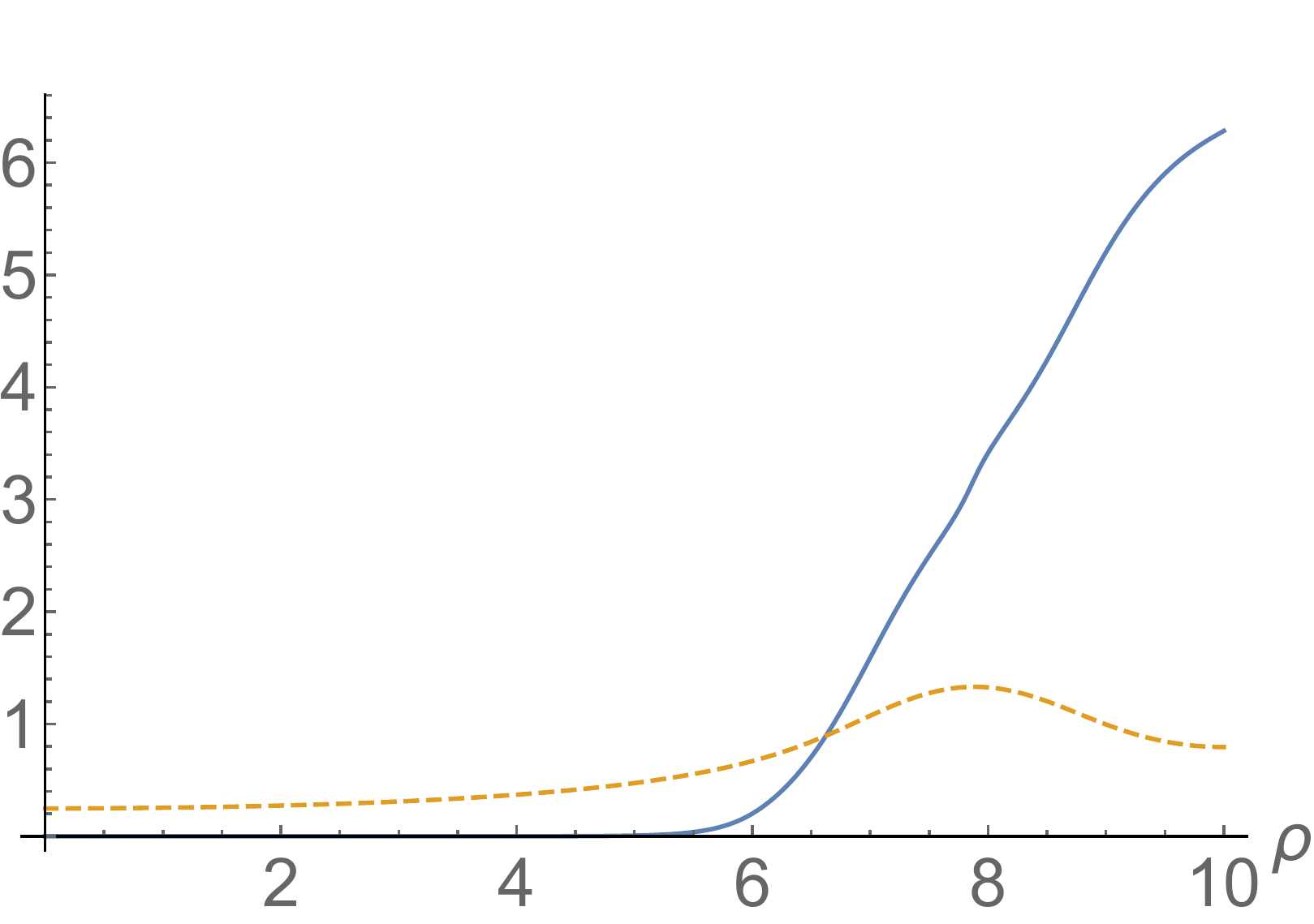}
\caption{}
\end{subfigure}
\begin{subfigure}{.5\textwidth}
\centering
\includegraphics[width=0.8\linewidth]{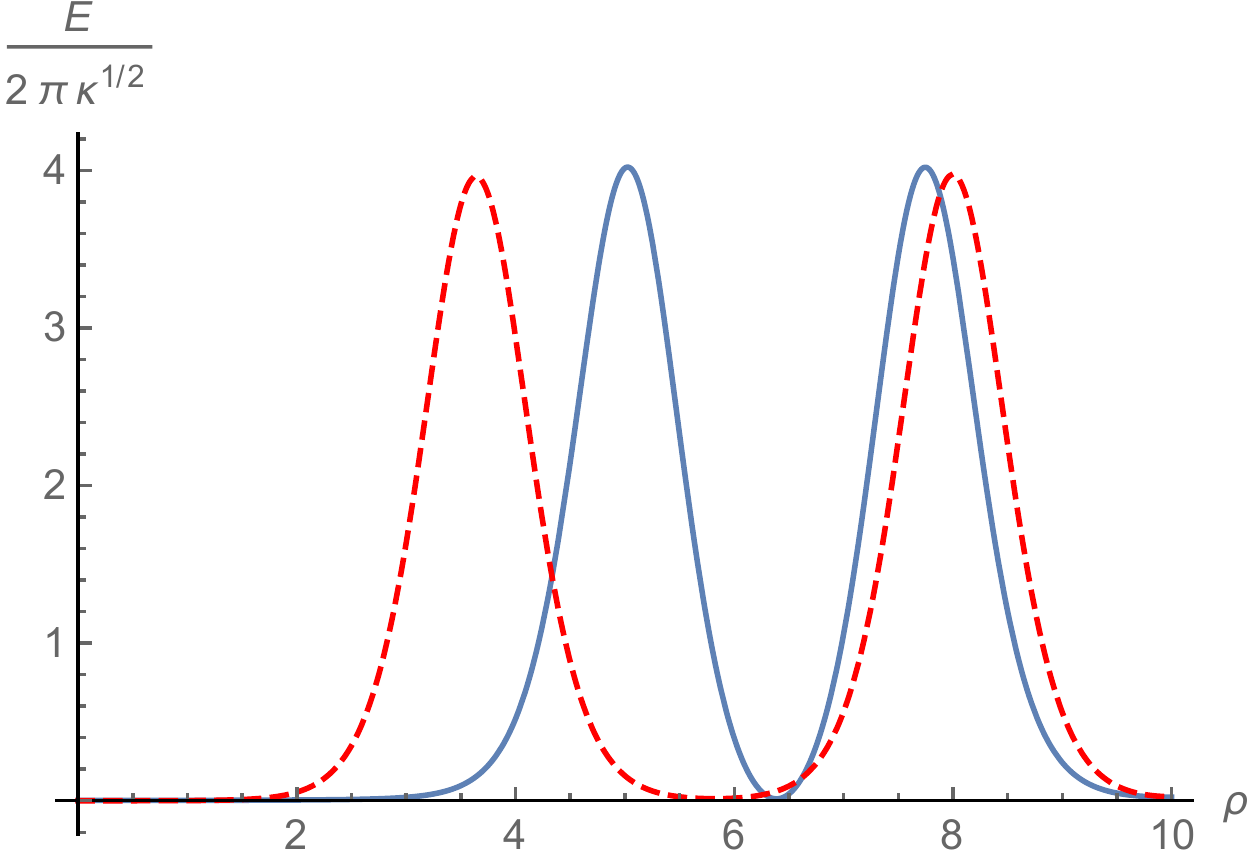}
\caption{}
\end{subfigure}
\begin{subfigure}{.5\textwidth}
\centering
\includegraphics[width=0.8\linewidth]{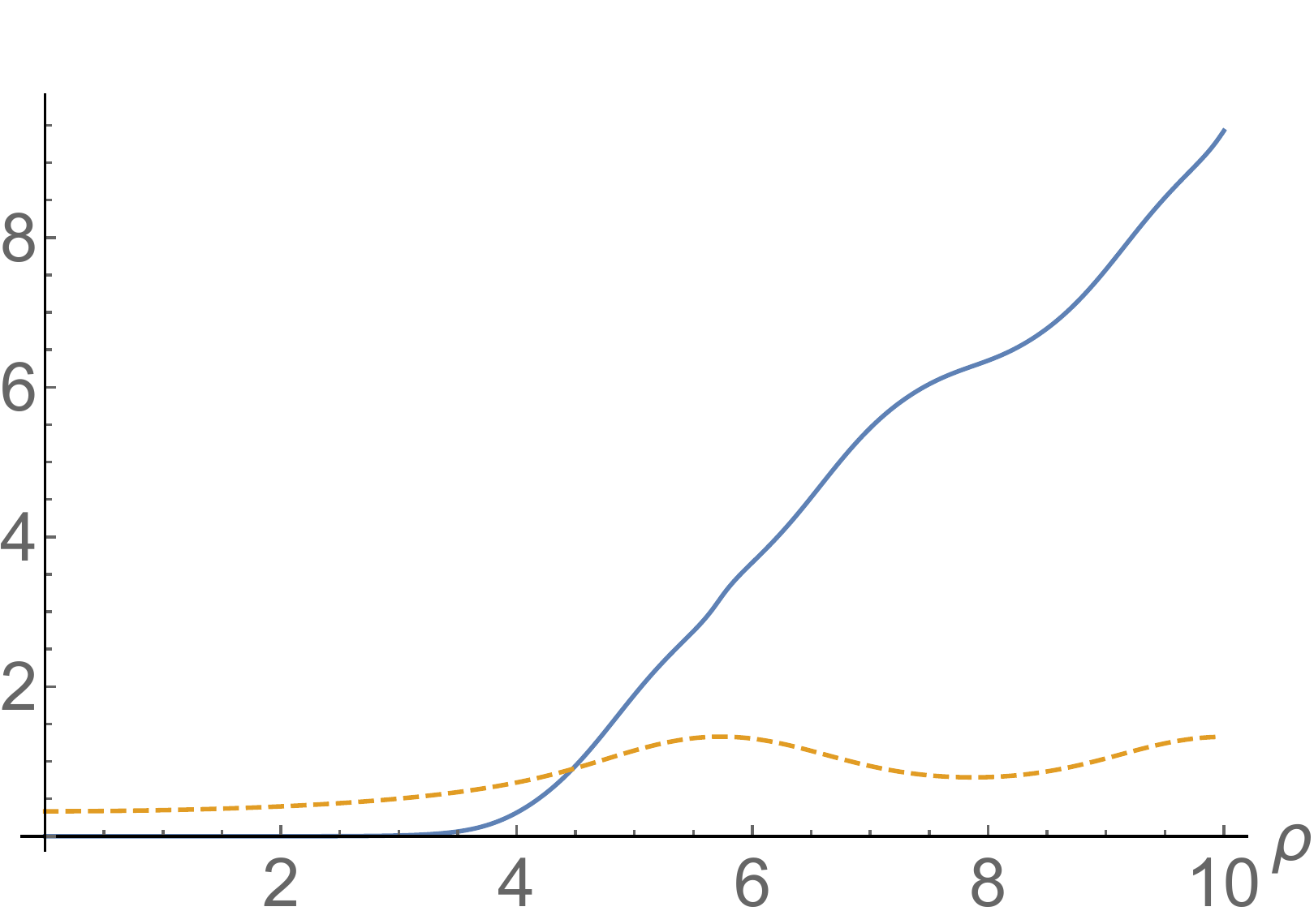}
\caption{}
\end{subfigure}
\begin{subfigure}{.5\textwidth}
\centering
\includegraphics[width=0.8\linewidth]{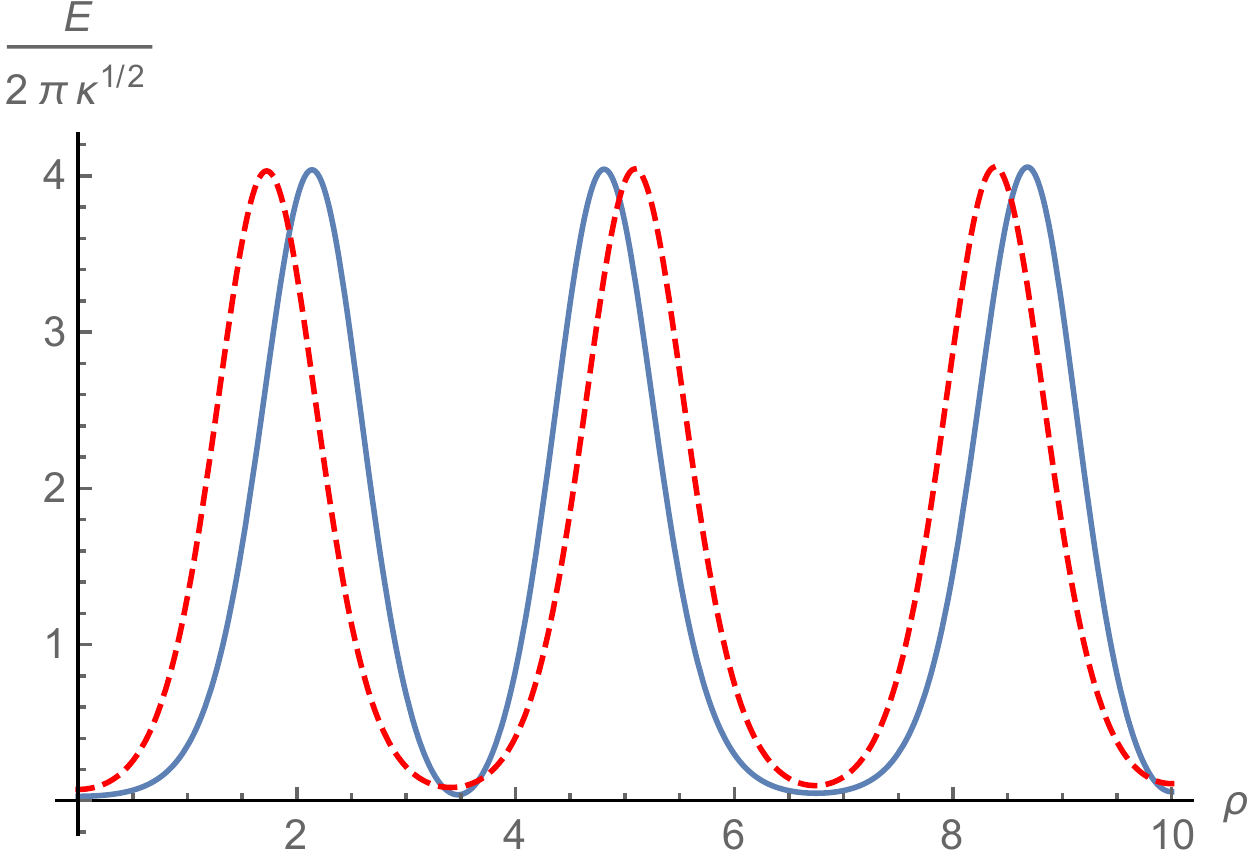}
\caption{}
\end{subfigure}
\begin{subfigure}{.5\textwidth}
\centering
\includegraphics[width=0.8\linewidth]{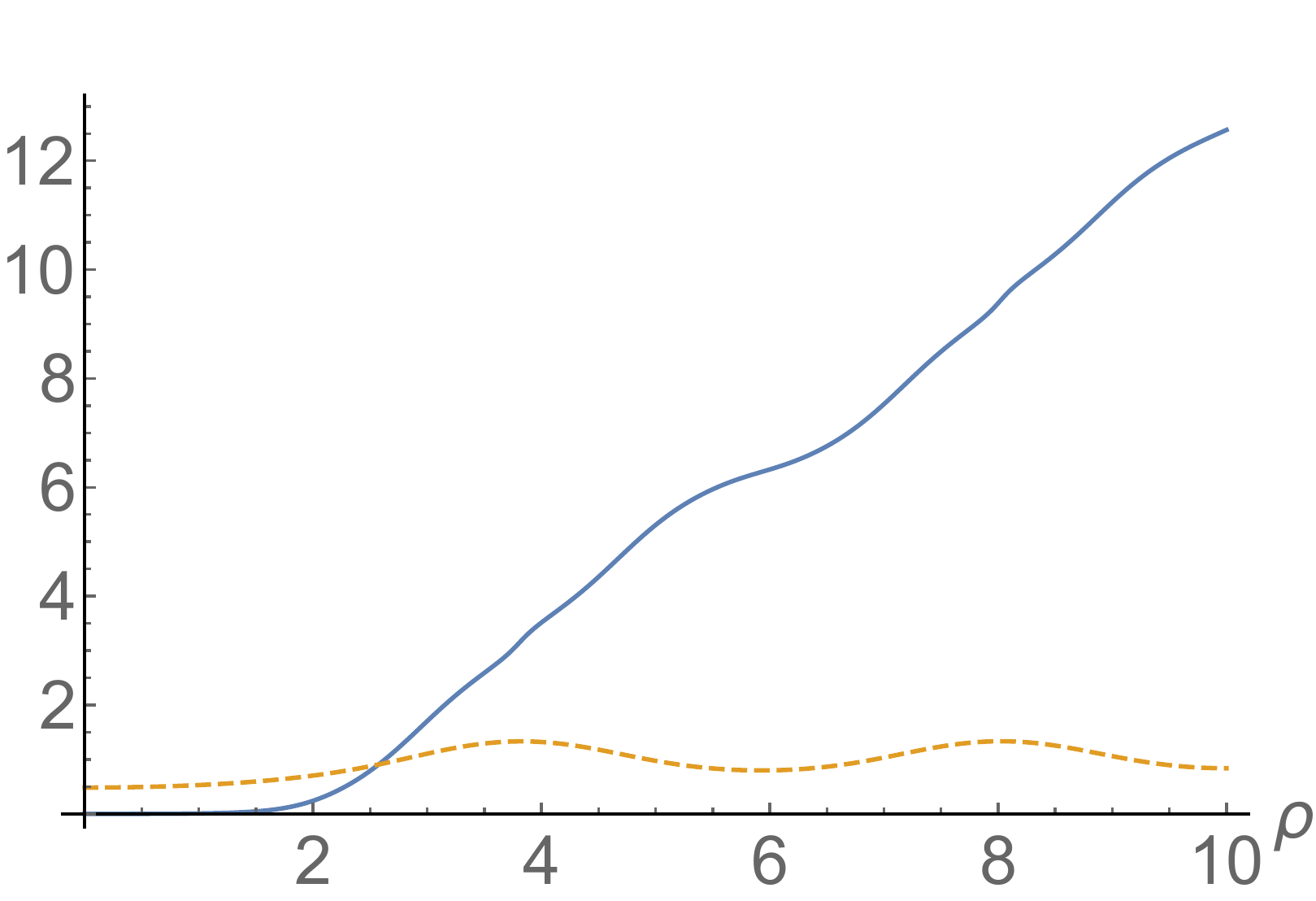}
\caption{}
\end{subfigure}
\begin{subfigure}{.5\textwidth}
\centering
\includegraphics[width=0.8\linewidth]{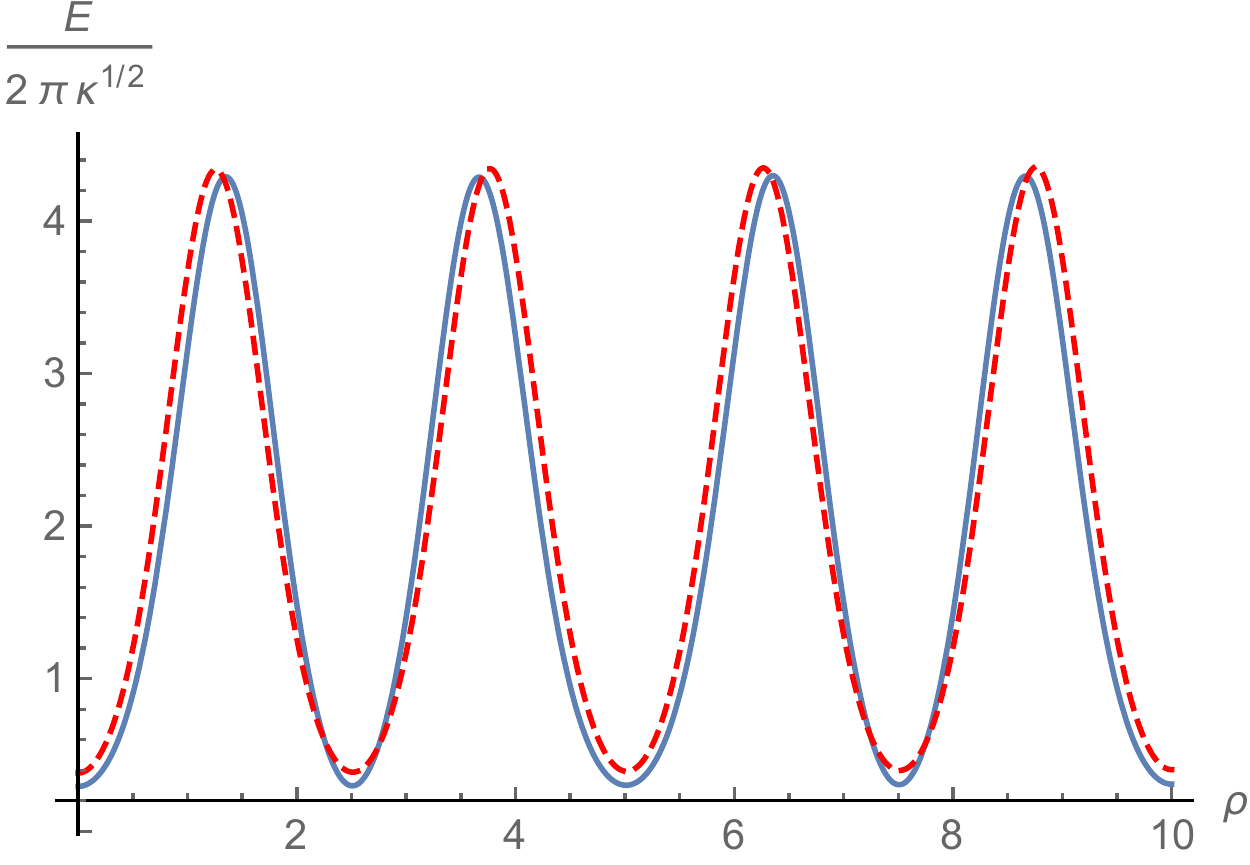}
\caption{}
\end{subfigure}
\begin{subfigure}{.5\textwidth}
\centering
\includegraphics[width=0.8\linewidth]{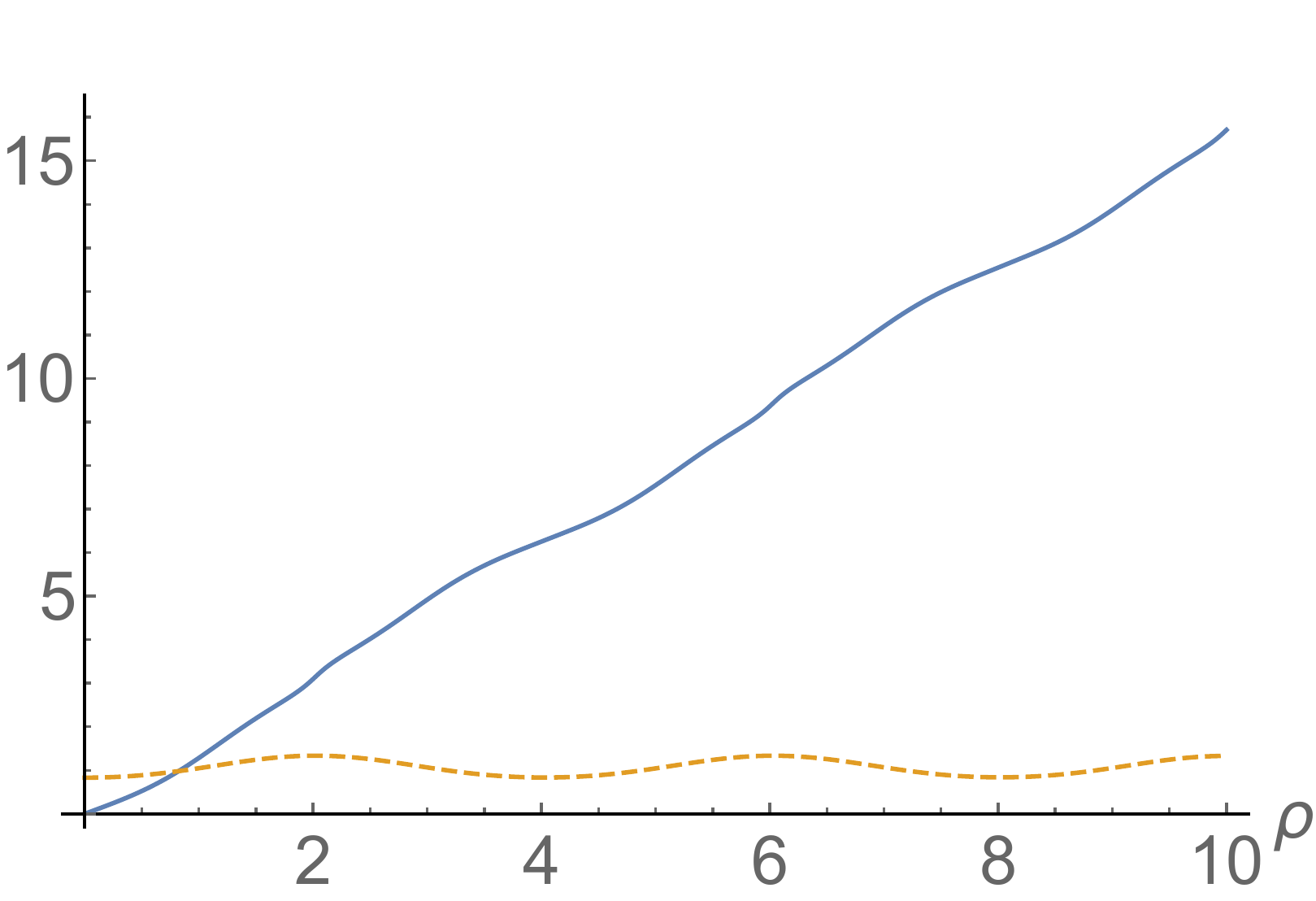}
\caption{}
\end{subfigure}
\begin{subfigure}{.5\textwidth}
\centering
\includegraphics[width=0.8\linewidth]{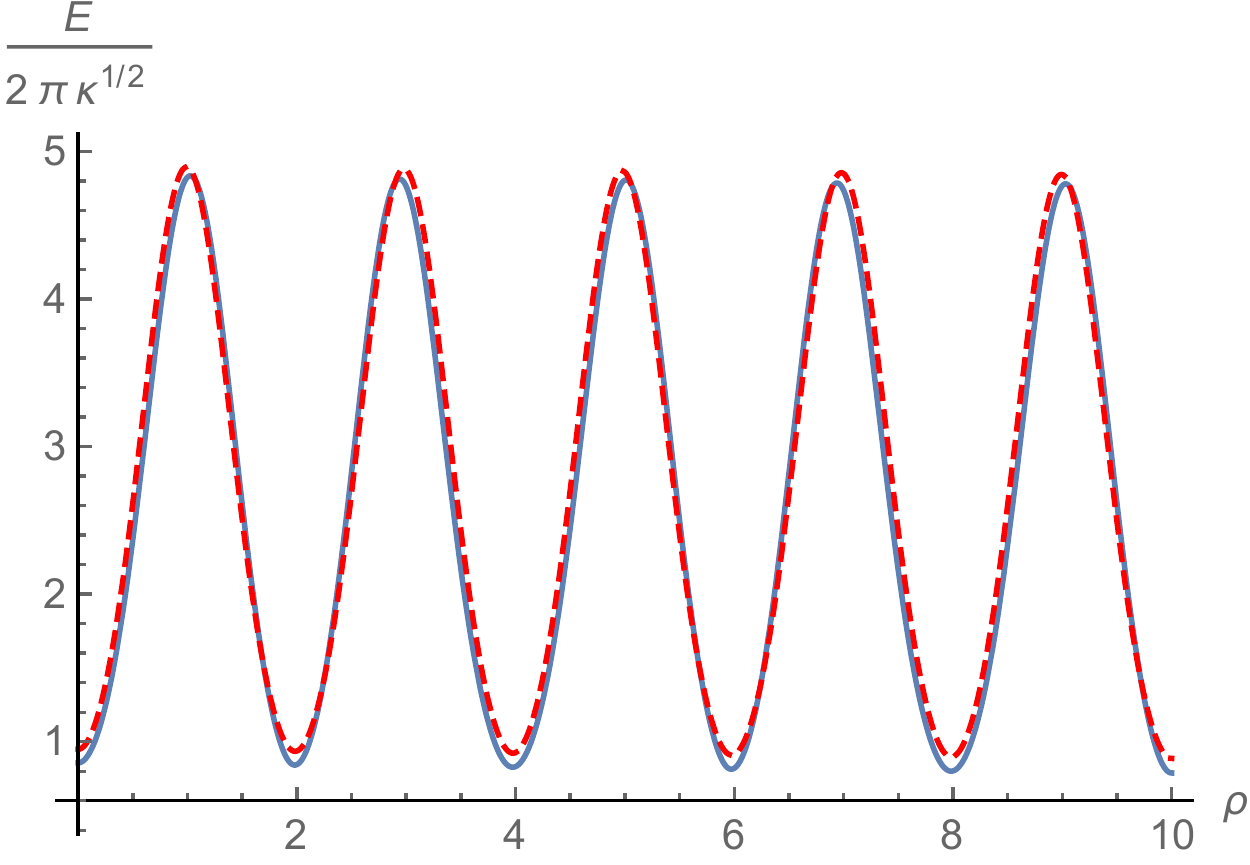}
\caption{}
\end{subfigure}
\caption{ All plots at $\protect\lambda = 1/8$, $|\protect\beta|=0.4$, $%
\Gamma=0$, $\protect\gamma/\protect\kappa =1/4$ and $\protect\sqrt{\protect%
\kappa}R=0.27$. The solid line is $\protect\alpha$ and the dashed line is $%
\protect\chi$. In the energy plots we include in dashed red the plots for
the energy of the Skyrmion solutions without any additional moduli. }
\label{fig11}
\end{figure}

\begin{figure}[ptb]
\centering
\includegraphics[width=0.7\linewidth]{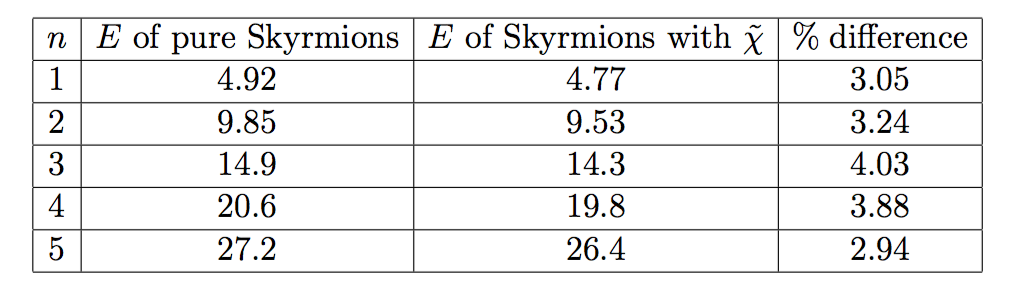}
\caption{}
\label{table}
\end{figure}

The solution corresponding to a single Skyrmion inside the tube is shown in
figure \ref{fig12}, along side its energy compared to the case without any
moduli. The multi-Skyrmion solutions are shown in figure \ref{fig11}. In
figure \ref{table} we calculate the dimensionless energies of the
corresponding solutions with their percentage differences defined by 
\begin{equation}
\%\; \text{diff} = \frac{E_{sk} - E_{\chi}}{E_{sk}}\times 100.
\end{equation}

We see that in finite volume, and in this particular topology, solutions
with orientational moduli are energetically preferred over solutions
without. This is in sharp contrast to our findings in flat space (described
in section 2). For fixed length $L$ of the tube, this difference grows with
the number of Skyrmions up until $n=3$, at which point it seems to drop. We find using our numerical procedure that there is a narrow range of parameters for which, for the same values of the parameters, convergence is seen up to $n=5$. Therefore, the results shown in this table are specific to a window of parameters with negligible variation from those quoted and may not necessarily follow a similar pattern generically. 
\newline

Furthermore, for fixed $L$, as seen by the energy plots, the presence of
additional orientational moduli serves to decrease the Skyrmion repulsion,
decreasing the separation between each Skyrmion (see figure \ref{fig11}
(b)). This effect is not a-priori surprising. The presence of an additional
bosonic scalar degree of freedom within the tube provides an attractive
force between Skyrmions which contrasts their repulsion. The $\chi$ field is
in this way partially screening the repulsive force.

The decrease of the repulsion energy would be even larger if one would
consider moduli which can condense separately on each elementary Skyrmion
(instead of being rigid, as implied by the ansatz in eq. (\ref{spaccimma2}%
)). However this would complicate the numerical analysis considerably as the
present system of coupled ODEs would become a system of non-linear coupled
PDEs (see discussion below). We hope to come back on this issue in a future
publication.

\begin{comment}
\begin{equation}\label{table}
\begin{array}{|c|c|c|c|}
\hline
n & \text{$E_{sk}$ of pure Skyrmions} & \text{$E_{\chi}$ of Skyrmions with }\tilde{\chi}
& \%\;\text{difference} \\ \hline
1 & 4.92 & 4.77 & 3.05 \\ \hline
2 & 9.85 & 9.53 & 3.24 \\ \hline
3 & 14.9 & 14.3 & 4.03 \\ \hline
4 & 20.6 & 19.8 & 3.88 \\ \hline
5 & 27.2 & 26.4 & 2.94 \\ \hline
\end{array}%
\end{equation}
\end{comment}

\subsection{Low energy theory of orientational moduli}

The low energy theory of the additional orientational moduli follows closely
that derived in section \ref{sec3}. Even though the solutions represent
multi-Skyrmions there still remains one explicit zero mode corresponding to
the equal global rotation of all the Skyrmions inside the global group. This
is seen precisely as per the flat case, taking $\tilde{\chi}$ as 
\begin{equation}
\chi^i=\tilde{\chi}(\rho )S^{i}(t),
\end{equation}%
with $S^{i}$ a unit vector satisfying $S^{i}S^{i}=1$. Then inserting this
parametrisation into eq.(\ref{action}) we obtain the low energy effective
action 
\begin{equation}
S_{le}=\frac{I_{2}}{2}\int dt\dot{S}^{i}\dot{S}^{i},\quad S^{i}S^{i}=1.
\end{equation}%
The constant $I_{2}$ now evaluates to 
\begin{equation}
\frac{I_{2}\sqrt{\kappa }}{4\pi }=\kappa R^{2}\int_{0}^{L}\tilde{\chi}%
^{2}d\rho =1.06.
\end{equation}

The presence of the extra moduli tends to decrease the repulsive
interactions between the elementary Skyrmions, hence it is natural to wonder
whether one can decrease this even further with a less rigid moduli
configuration. The natural guess is that if the moduli could condense on
each elementary Skyrmion independently, then the decrease of the repulsive
interactions would be even larger. However, an ansatz of the form in eq. (%
\ref{spaccimma2}) is not suitable to achieve this goal. The reason is that
the factorized expression $\chi^i=\tilde{\chi}(\rho )S^{i}(t)$\ implies that
the extra isospin number related to the unit vector $S^{i}(t)$\ is the same
in any point of the tube. We would like to describe a situation in which the
orientation of the unit vector $S^{i}(t)$\ depends on the position of the
tube and, in particular, is able to distinguish the elementary Skyrmions. In
this respect, a reasonable ansatz is%
\begin{equation*}
\chi^i_{L}=\tilde{\chi}(\rho )S^{i}(t,\rho )\ ,
\end{equation*}%
in order to be able to have a different $S^{i}$ on each elementary Skyrmion
living within such a finite volume region. However, such an ansatz would
lead to a system of coupled PDEs and would complicate the numerical
analysis. We hope to come back on this interesting issue in a future
publication.

We can however make some intuitive progress by considering a dramatic
simplification. If we assume that the $\tilde{\chi}$ field is approximately
localised on each Skyrmion in the tube then we may consider an ansatz of the
form 
\begin{equation}
\chi (\rho )^{i}\approx \sum_{j=1}^{n}\chi _{j}\Pi (\alpha _{j}\delta (\rho
-\rho _{j}))S_{\rho _{j}}^{i}(t),
\end{equation}%
where $\chi _{j}\Pi (\alpha _{j}\delta (\rho -\rho _{j}))$ is a rectangular
function of width $\alpha _{j}$ and height $\chi _{j}$ localised on each
Skyrmion centre $\rho _{j}$, to which we assign an orientational moduli
vector $S_{\rho _{j}}^{i}(t)$. The ansatz represents a total function $\chi
(\rho )^{i}$ in the length of the tube as the sum of each individual
Skyrmion contribution, crudely approximated by a step function. In this case
we can immediately observe how moduli interactions arise, consider for
example the potential term 
\begin{equation}
\chi ^{i}\chi ^{i}\approx \sum_{k=1}^{n}\sum_{j=1}^{n}\chi _{j}\chi
_{k}S_{\rho _{j}}^{i}(t)S_{\rho _{k}}^{i}(t)\Pi (\alpha _{j}\delta (\rho
-\rho _{j}))\Pi (\alpha _{k}\delta (\rho -\rho _{k})),
\end{equation}

then in the limit in which the rectangular functions have infinitesimal
width $|\rho_i-\rho_j| >> |\alpha_i-\alpha_j|$ (the step function tend to
delta functions), so that each Skyrmion modulus is perfectly localised, this
expression is independent of the moduli since the condition $S^i_j S^i_j =1$
applies (no sum over $j$). However, if the moduli overlap within a finite
region, $|\rho_i-\rho_j| \approx |\alpha_i-\alpha_j|$, then terms with $%
S^i_j S^i_k $ with $j\neq k$ are non-vanishing. These kind of terms are
clearly present in the solutions we observe numerically in figure \ref{fig11}
where $\tilde{\chi}(\rho)$ is dispersed and no-where vanishing within the
tube. Through these interactions the moduli are lifted and become
quasi-moduli. The only remaining true modulus is the global rotation
discussed at the beginning of this section.

\subsection{Large $R$ flat limit}

As already remarked, the technical advantage of the geometry in eq. (\ref%
{spaccimma3}) lies in the fact that it allows one to keep the symmetries of
the hedgehog ansatz without losing information about the finiteness of the
volume where the elementary Skyrmions live. Per se this provides a very
simple framework to analyze the issue of (orientational) moduli in the
presence of configurations of arbitrary Baryon number. Such tube-shaped
regions are however not flat as the curvature of this metric is proportional
to $\frac{1}{R^{2}}$. \newline

\begin{figure}[ptb]
\begin{subfigure}{.5\textwidth}
\centering
\includegraphics[width=0.8\linewidth]{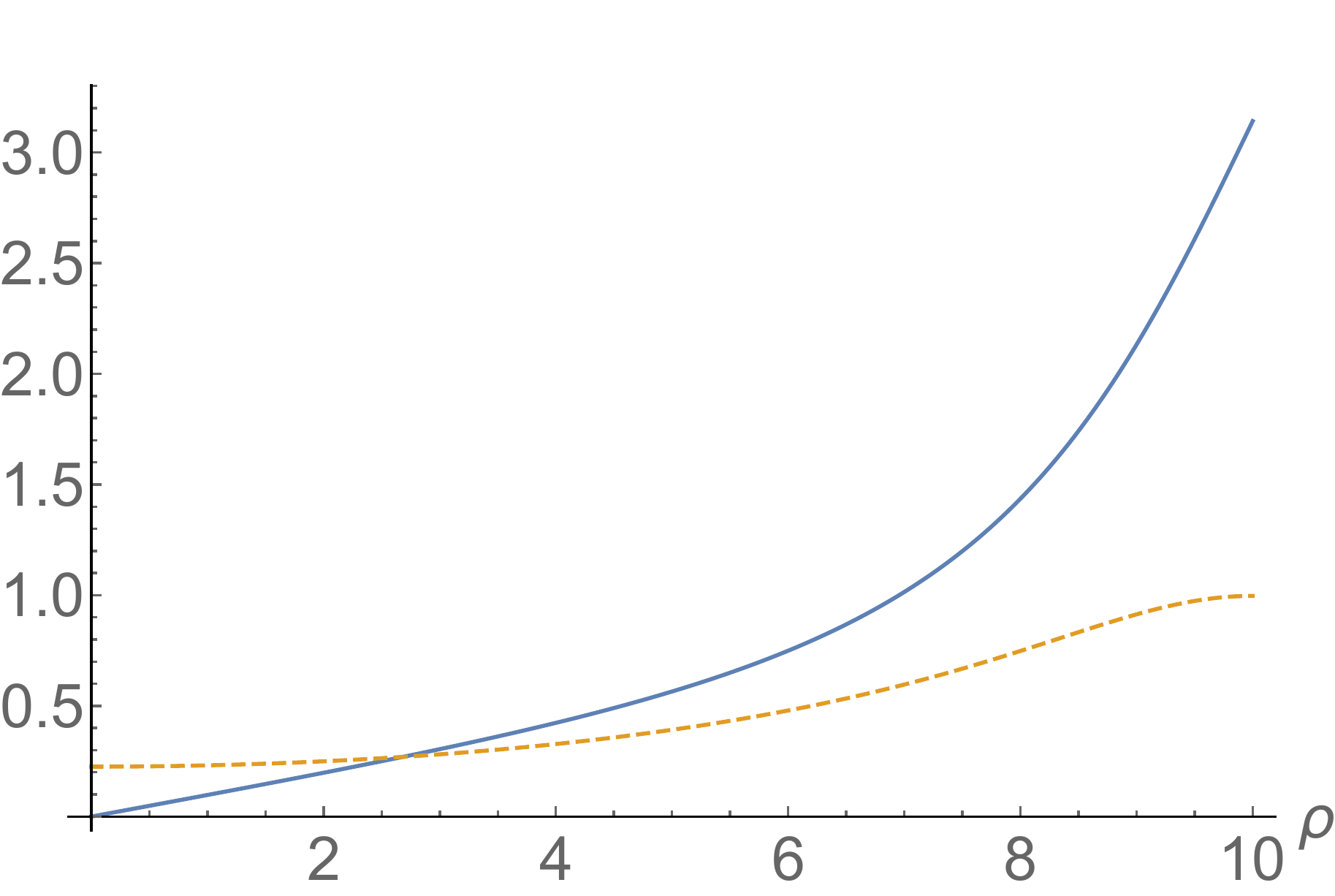}
\caption{}
\end{subfigure}
\begin{subfigure}{.5\textwidth}
\centering
\includegraphics[width=0.8\linewidth]{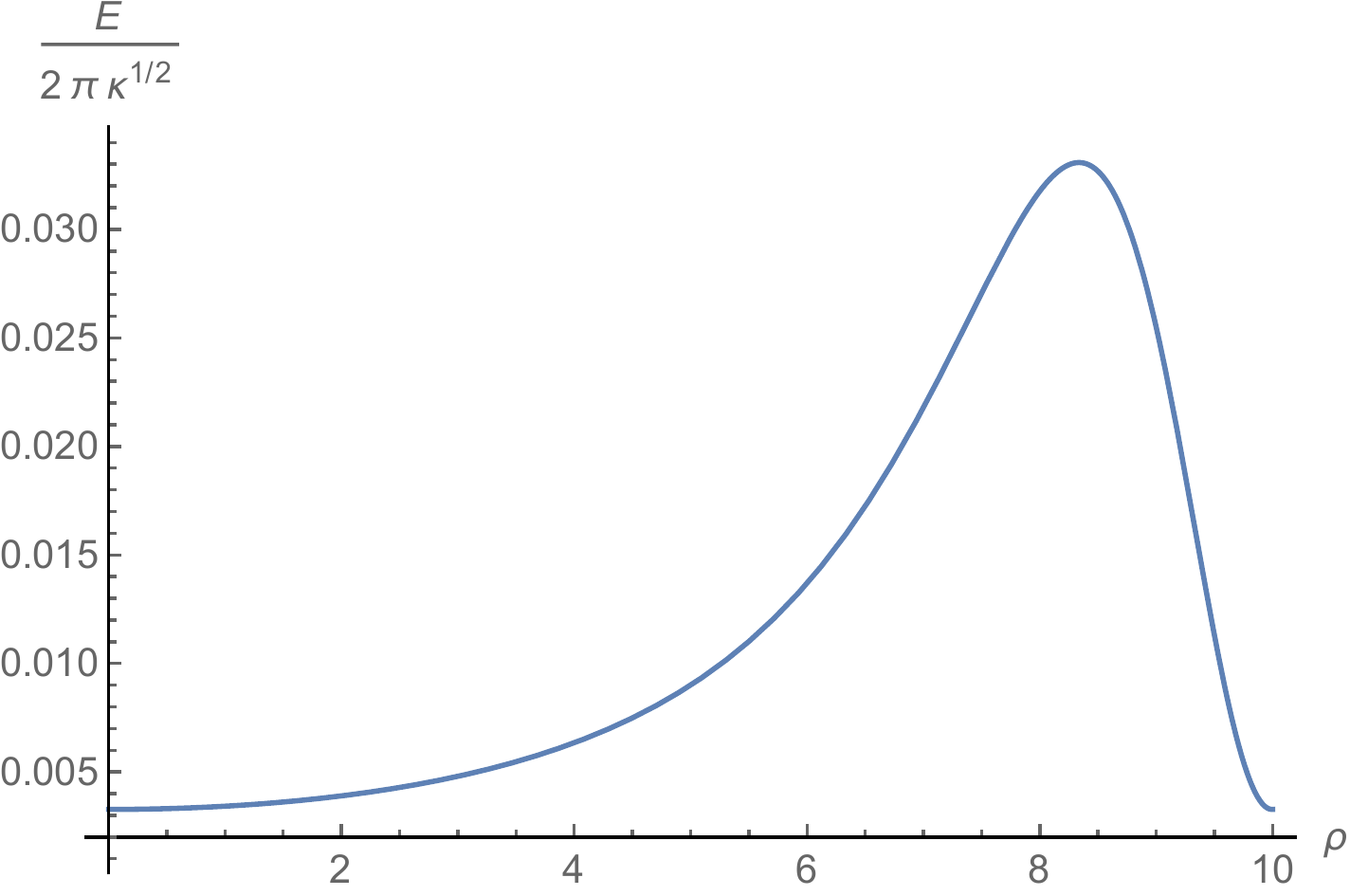}
\caption{}
\end{subfigure}
\begin{subfigure}{.5\textwidth}
\centering
\includegraphics[width=0.8\linewidth]{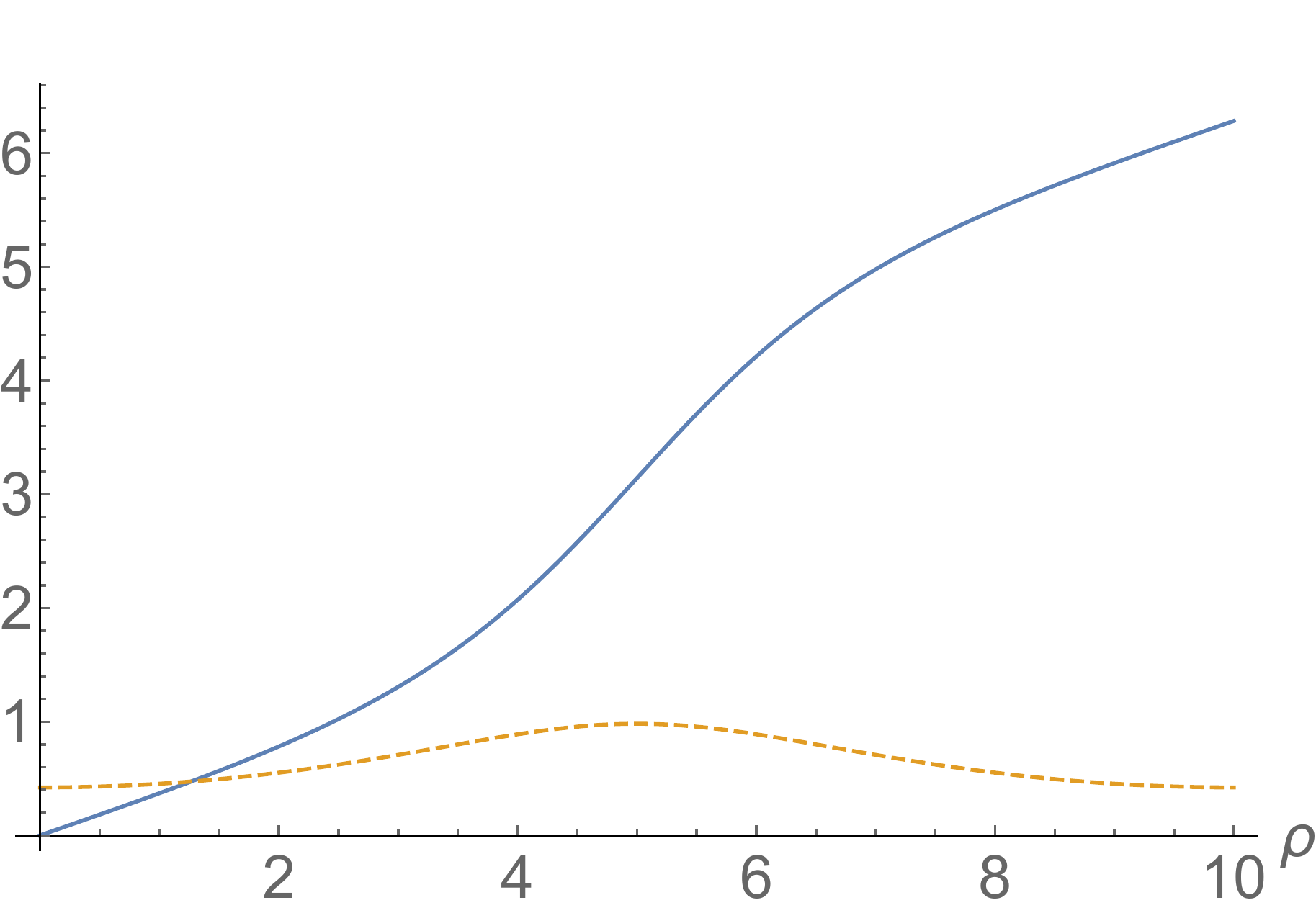}
\caption{}
\end{subfigure}
\begin{subfigure}{.5\textwidth}
\centering
\includegraphics[width=0.8\linewidth]{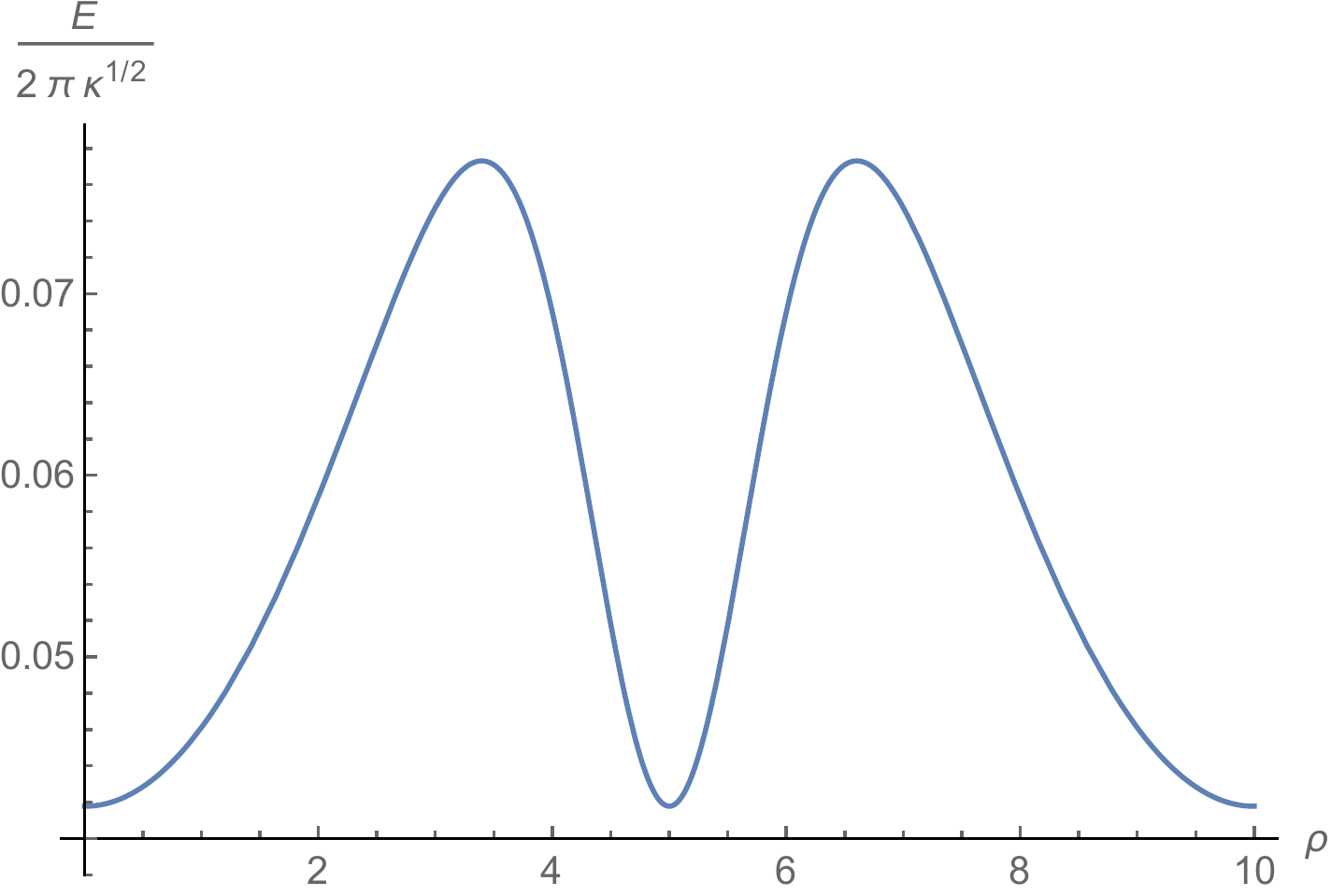}
\caption{}
\end{subfigure}
\begin{subfigure}{.5\textwidth}
\centering
\includegraphics[width=0.8\linewidth]{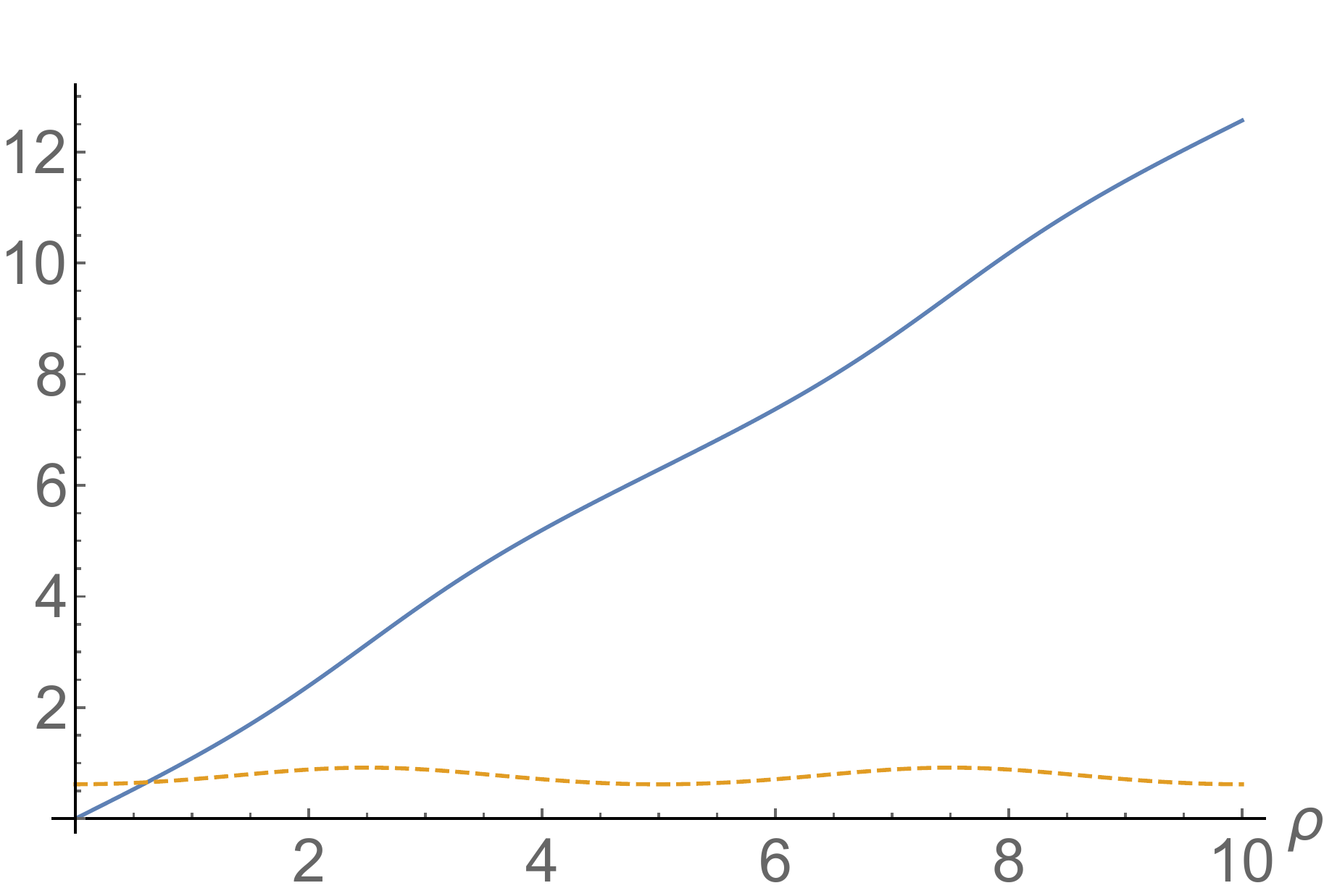}
\caption{}
\end{subfigure}
\begin{subfigure}{.5\textwidth}
\centering
\includegraphics[width=0.8\linewidth]{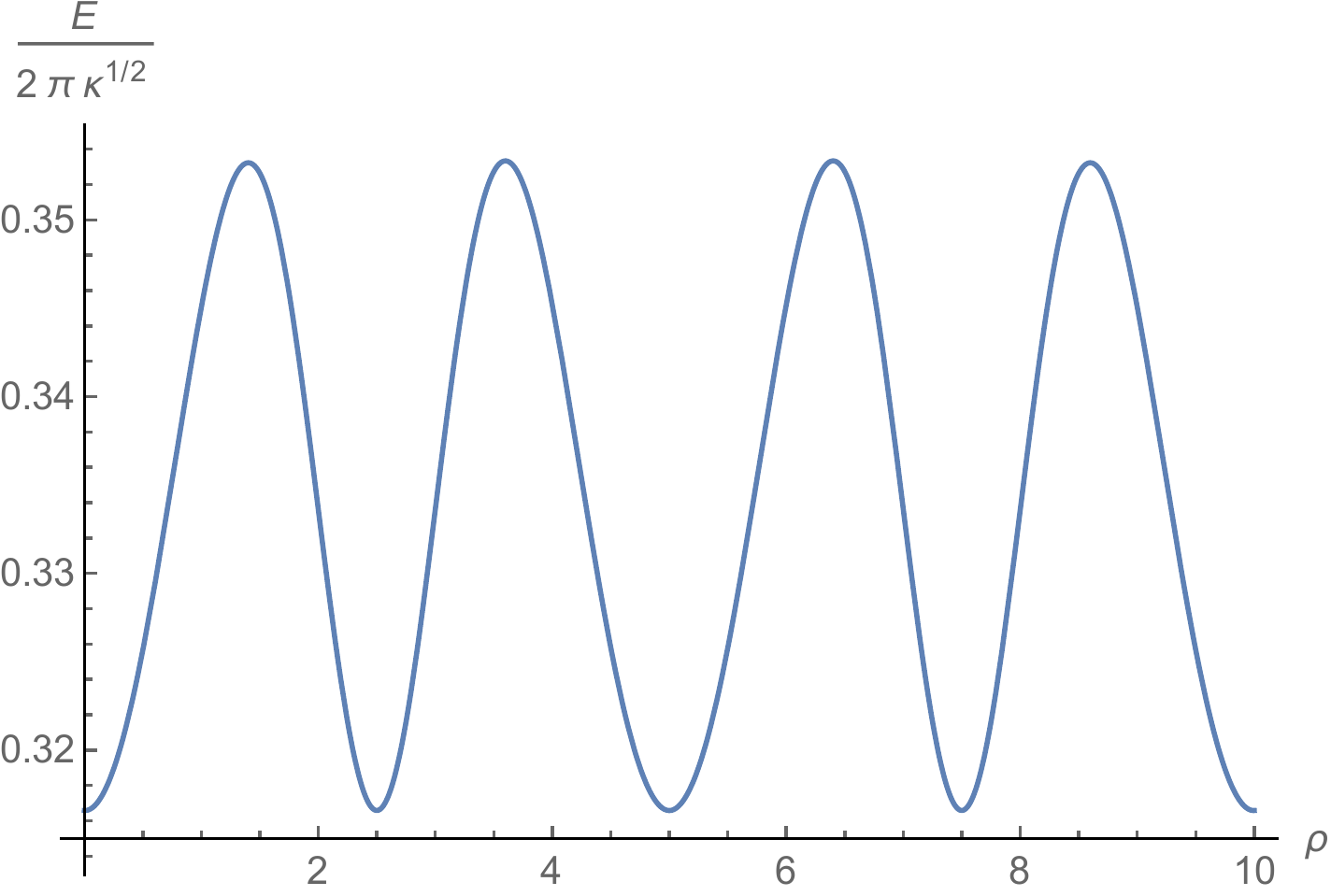}
\caption{}
\end{subfigure}
\caption{ All plots at $\protect\lambda = 1/8$, $|\protect\beta|=0.4$, $%
\Gamma=0$, $\protect\gamma/\protect\kappa =1/4$. The solid line is $\protect%
\alpha$ and the dashed line is $\protect\chi$. }
\label{figflat}
\end{figure}

As far as the local effects of the curvature are concerned, they can be
considered small when they are negligible with respect to the corrections
which the Skyrme model receives in the large $\mathbf{N}$ expansion of QCD 
\cite{49}. This happens already when $R\sim 100\ fm$. However, one can take
a formal large $\sqrt{\kappa}R$ limit directly in the field equations (\ref%
{eqR1})-(\ref{eqR2}) which, in the leading order of such expansion, read

\begin{equation}
\alpha ^{\prime \prime }-\frac{4\gamma }{\kappa }(\sin \alpha )\tilde{\chi}%
^{2}=0,
\end{equation}%
\begin{equation}
\frac{\gamma }{\kappa }\left( 4(-1+\cos \alpha )+\Gamma \right) \tilde{\chi}%
+2|\beta |\tilde{\chi}^{3}-\tilde{\chi}^{\prime \prime }=0.
\end{equation}%
The above system is considerably simpler but still non-trivial due to the
non-linear interactions which are still present. Solving the above equations
with boundary conditions of arbitrarily Baryon number (those of eqs (\ref%
{bc1}) and (\ref{bc2})) can be though to represent, in the flat case,
multi-Skyrmions configurations constrained to live within flat tubes whose
sections have dimension much bigger than $1$ $fm$. Each elementary Skyrmion
belonging to these multi-Skyrmionic configurations is very well localized in
the direction of the axis of the tube (namely, the energy density profile in
the large $R$ limit looks almost like the superposition of many
non-overlapping peaks, one for each elementary Skyrmion). On the other hand,
in the spatial directions orthogonal to the axis of the tube, the Skyrmions
are homogeneous (in other words, the energy density profile does not depend
on the coordinates transverse to the axis). Solutions of this form are shown
in figure \ref{figflat}. Note that in this case the solution with $\tilde{%
\chi}=0$ is just the linear function $\alpha =\frac{n\pi }{L}\rho $ which
reduces the energy to an $n$ dependent constant. \newline

An interesting issue on which we hope to come back in a future publication
is to develop in a systematic way the large $R$ expansion in which (a
suitable "adimensionalized" version of) $\frac{1}{R^{2}}$ would play the
role of small parameter.

\section{Conclusions}

In the present paper, combining the construction of analytic
multi-Skyrmionic configurations with the recent approach to the analysis of
orientational moduli \cite{shifman1}, we analyzed how extra orientational
moduli affect the properties of multi-Skyrmionic configurations of the
four-dimensional Skyrme model. This analysis sheds light on the peculiar
behavior of orientational moduli when multi-solitonic configurations are
present. It reveals interesting novel features. First of all, when
considering finite geometry, the orientational moduli tend to decrease the
repulsive interactions among elementary $SU(2)$ Skyrmions (however, this
effect decreases with the increase of the Baryon number). Moreover, in the
case of a single Skyrmion, the appearance of moduli is energetically
favorable if finite volume effects are present otherwise, in the usual flat
topologically trivial case, it is not.

\section{Acknowledgements}

This work has been funded by Fondecyt grants 1160137 and 3140122. The Centro
de Estudios Científicos (CECs) is funded by the Chilean Government through
the Centers of Excellence Base Financing Program of Conicyt.

\end{document}